\documentclass[11pt,twocolumn,aps,prl,reprint]{revtex4-1}

\usepackage{braket}

\usepackage{graphicx,amsmath,amssymb,braket}

\begin{document}

\title{Emission and amplification of surface plasmons in resonant - tunneling van der Waals heterostructures}
\author{D. Svintsov$^{1}$, Zh. Devizorova$^{2,3}$, T. Otsuji$^{4}$, and V. Ryzhii$^{4}$}
\affiliation{$^{1}$Laboratory of Nanooptics and Plasmonics, Moscow Institute of Physics and Technology, Dolgoprudny 141700, Russia}
\affiliation{$^{2}$Department of Physical and Quantum Electronics, Moscow Institute of Physics and Technology, Dolgoprudny 141700, Russia}
\affiliation{$^{3}$Kotelnikov Institute of Radio Engineering and Electronics, Russian Academy of Science, Moscow, 125009 Russia}
\affiliation{$^{4}$Research Institute of Electrical Communication, Tohoku University,  Sendai 980-8577, Japan}

\email{svintcov.da@mipt.ru}
 
\begin{abstract}

We predict a new mechanism of surface plasmon amplification in graphene-insulator-graphene van der Waals heterostructures. The amplification occurs upon the stimulated interlayer electron tunneling accompanied by the emission of a coherent plasmon. The quantum-mechanical calculations of the non-local high-frequency tunnel conductivity show that a relative smallness of the tunneling exponent can be  compensated by a strong resonance due to the enhanced tunneling between electron states with collinear momenta in the neighboring graphene layers. With the optimal selection of the barrier layer, the surface plasmon gain due to the inelastic tunneling can  compensate or even exceed  the loss due to both Drude and interband absorption. The tunneling emission of the surface plasmons is robust against a slight twist of the graphene layers and might explain the  electroluminescence from the tunnel-coupled graphene layers observed in the recent experiments.
\end{abstract}

\maketitle

The ultrarelativistic nature of electrons in graphene gives rise to the uncommon properties of their collective excitations -- surface plasmons~\cite{Graphene_plasmonics-1,Graphene_plasmonics-2,Graphene_plasmonics-3}. The deep subwavelength confinement~\cite{Graphene_plasmonics-3}, the unconventional density dependence of frequency~\cite{Ryzhii-plasmons,Das_Sarma_Plasmons}, and the absence of Landau damping~\cite{Ryzhii-plasmons} are probably the most well-known features of  plasmons in graphene-based heterostructures.  Among more sophisticated predictions there stand the existence of weakly damped transverse electric plasmons~\cite{Mikhailov_new_mode} and quasi-neutral electron-hole sound waves near the neutrality point~\cite{Our-hydrodynamic, New_plasmon_mode}. Some peculiar types of plasmons can be excited in the graphene $p-n$ junctions~\cite{Plasmons_pn}, field- effect transistors~\cite{Polini_FET, Our_NLHD}, optoelectronic modulators~\cite{Plasma-resonances-in-modulator}, and nanomechanical resonators~\cite{Our_NEMS} engaging for the improved device performance at the terahertz frequencies.

Unfortunately, the experimental studies of graphene plasmons are yet unable to confirm or refute many of these predictions. To achieve an extreme plasmon confinement, one has to sacrifice their propagation length. The latter is of the order of several micrometers at the infrared frequencies~\cite{Fei_nano_imaging,Koppens_nano_imaging} and is limited by the interband absorption in intrinsic samples and somewhat lower Drude absorption in the doped ones~\cite{Principi_plasmon_loss_hBN}. The experimentally reported quality factors of graphene plasmons reach only five for graphene on SiO$_2$~\cite{Fei_nano_imaging,Koppens_nano_imaging}, and 25 for graphene encapsulated in hexagonal boron nitride~\cite{Koppens_nano_imaging_hBN} at room temperature. In the latter case, the damping is due to the scattering by the intrinsic acoustic phonons~\cite{Principi_plasmon_loss_hBN} and can be suppressed only by lowering the temperature.

Instead of reducing the plasmon loss, it is possible to overcome the damping by introducing the gain medium which can replenish the energy being dissipated upon scattering. This idea has stimulated the re-examination of various 'classical' plasma instabilities in graphene, including the beam and resistive instabilities~\cite{Mikhailov-THz}, Dyakonov-Shur self-excitation~\cite{Polini_FET, Our_NLHD}, and generation due to the negative differential conductance~\cite{Berardi_APL}. On the other hand, the plasmon gain can be provided by the photogenerated electrons and holes recombining with plasmon emission~\cite{Dubinov_JPCM}, which opens up the prospects of graphene-based spasers~\cite{Rana_IEEE}. However, relatively fast nonradiative recombination in graphene~\cite{Winnerl_PRL,Gierz_JPCM} hinders the practical implementation of those structures.

In this paper, we study the inelastic tunneling in graphene-dielectric-graphene van der Waals heterostructures accompanied by the emission of surface plasmons (SPs), and the possibility of the coherent SP amplification due to the tunneling gain. The inelastic plasmon-assisted tunneling was first observed in the late 70's; it was shown to be responsible for the light emission from metal-insulator-metal tunneling diodes~\cite{Lambe_PRL}. Afterwards, the tunneling excitation of SPs was demonstrated using the scanning probes above the metal surfaces~\cite{Berndt-inelastic-tunneling-plasmons,Persson-theory-tip-plasmon,Novotny-tip-plasmon}. However, the possibility of the SP gain due to the tunneling to exceed the loss has been never considered realistic due to the smallness of the tunnel exponent, thus all experiments to date have reported only on the spontaneous SP emission. To achieve large tunneling gain, one needs some resonant feature to compensate the smallness of the tunnel exponent. As an example, such a resonance occurs in the quantum cascade lasers due to the alignment of energy subbands in the neighboring quantum wells~\cite{Kazarinov_Suris,Capasso_science,Ryzhii_DGL_laser}. In this paper, we show that the frequency dependence of plasmon tunneling gain in double graphene layer structures also exhibits a strong resonance due to the enhanced interaction between the Dirac electrons having collinear momenta~\cite{Fritz_PRB}. In twisted graphene layers, the finite momentum of emitted plasmon can bridge the electron dispersions in the neighboring layers together~\cite{Twist-controlled}, which makes the effect of plasmon tunneling emission robust against a slight layer twisting. We also discuss the recent experimental observations of the terahertz emission from the double graphene layer heterostructures~\cite{THz_emission_DGL_experim} and address the role of the inelastic-tunneling plasmon excitation in the observed spectra.

\begin{figure}[ht]
\includegraphics[width=0.9\linewidth]{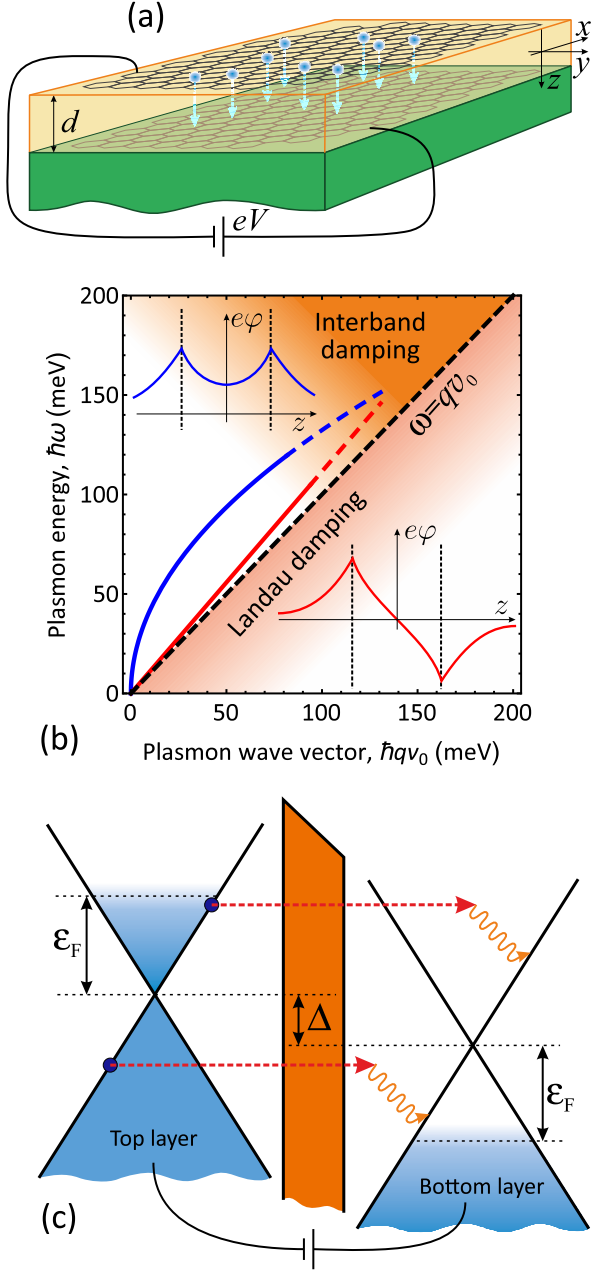}
\caption{\label{Structure} (a) Schematic views  of the double-graphene-layer tunnel heterostructure. (b) The spectra of the supported plasmon modes calculated for $d = 3 nm$, $\varepsilon_F = 100$ meV, $T=300$ K. Insets in show the spatial profiles of the plasmon potential in the optical (top) and acoustic (bottom) modes. (c) The band diagram of the double graphene layer hetero-structure. The straight red arrows correspond to electron tunneling between graphene layers accompanied by the emission of surface plasmons (wavy arrows).
}
\end{figure}

The double-graphene layer heterostructures shown schematically in Fig.~1 a support optical and acoustic surface plasmon modes having symmetric and anti-symmetric distributions of electric potential\cite{Hwang_PRB_2GL,Voltage_controlled}.  Their spectra are studied in detail in the absence of tunneling: the dispersion of symmetric (optical) mode is square-root, $\omega_+ \approx v_0 \sqrt{4 \alpha_c q q_F}$, where $v_0$ is the Fermi velocity in graphene, $q_F$ is the Fermi wave vector, $\alpha_c = e^2/\kappa \hbar v_0$ is the coupling constant, while the dispersion of the antisymmetric (acoustic) mode is sound-like, $\omega_- = s q $ (Fig. 1b). Its velocity approaches the Fermi velocity $v_0$ as the interlayer distance $d$ decreases, but newer falls below it into the region of Landau damping. In realistic tunnel-coupled double layers ($d\approx 3$ nm, $\varepsilon_F \approx 0.1$ eV), $s = 1.15 v_0$. Just the acoustic mode has a nonzero average field strength between the two layers and thus can induce interlayer tunneling~\cite{DasSarma-PRL-tunnel-plasmon}, a process in which an electron passes from one layer to another with emission of a coherent plasmon (Fig. 1 c). The tunneling can be included into the dispersion law of acoustic SPs by considering the tunnel current density $J_{\bot {\bf q}\omega}$ as a generation-recombination term in the continuity equations for electrons in a single layer~\cite{Ryzhii_Shur_JJAP}. In the linear response mode, this current is proportional to the interlayer potential difference, $J_{\bot {\bf q}\omega} = G_{{\bf q}\omega}(\varphi_{t,{\bf q}} - \varphi_{b,{\bf q}})$, where $G_{{\bf q}\omega}$ is the tunnel conductance. In these notations, the dispersion law of the acoustic SP mode becomes
\begin{equation}
\label{Acoustic_mode}
-\frac{i\omega \kappa }{4\pi q}\left( 1+\coth \frac{qd}{2} \right)+ \sigma_{\bf{q}\omega } + \frac{2{G_{\bf{q}\omega }}}{q^2}=0,
\end{equation}
where $\sigma_{\bf{q}\omega}$ is the in-plane conductivity of graphene. Its real part, ${\rm Re}\sigma_{\bf{q}\omega}$, is positive and leads to the SP damping. At the same time, ${\rm Re} G_{\bf{q}\omega}$ can be negative, at least, at low energy of SP quanta $\hbar \omega$. The negative tunnel conductance appears as the number of electrons with the energy $E$ in top layer exceeds the number of electrons with energy $E-\hbar \omega$ in the bottom layer; such carrier distribution can be viewed as a 'remote population inversion'. The sign of the quantity ${\rm Re}\left[ \sigma_{\bf{q}\omega} + 2 G_{\bf{q}\omega}/q^2 \right]$ thus determines whether the SPs propagating along the double-layer are being damped or amplified. 

An only missing ingredient to judge on the possibility of the net plasmon gain is the theory of  non-local ($q \neq 0$) high-frequency tunnel conductivity of van der Waals structures; the latter, to the best of our knowledge, has been studied only in the DC limit~\cite{Brey_PRA,Vasko_PRB}. To construct this theory, we start with the tight-binding Hamiltonian of the tunnel coupled layers in the presence of the propagating plasmon (we set $\hbar \equiv 1$):
\begin{equation}
\label{Hamiltonian}
\hat{H}=\left( \begin{matrix}
   {{{\hat{H}}}_{G+}} & {\hat{\mathcal{T}}}  \\
   {{{\hat{\mathcal{T}}}}^{*}} & {{{\hat{H}}}_{G-}}  \\
\end{matrix} \right) + {\hat H}_{\rm{int}} \equiv {\hat H}_0 + {\hat H}_{\rm{int}}.
\end{equation}
Here, ${\hat H}_{G\pm} = v_0 {\bf \sigma} {\hat {\bf p}} \pm {\hat I}\Delta/2$ describe separate graphene layers, $\Delta$ is the energy spacing between the Dirac points, $\hat{\bf p}$ is the in-plane momentum operator, $\hat I$ is the identity matrix, and $\hat{\mathcal T}$ is the tunneling matrix. For the sake of analytical traceability, we choose $\hat{\mathcal T}$ in its simplest form which is applicable for AA-stacking of aligned graphene layers~\cite{Brey_PRA,Twisted_GBL}, $\hat{\mathcal T} = \Omega {\hat{I}}$, where $\Omega$ is the 'tunneling frequency'. The interaction part, ${\hat H}_{\rm{int}}$, describes the electron-plasmon coupling; its matrix elements are calculated as the overlap of the eigen functions of $\hat H_0$ with the electron potential energy in the field of SP $e\varphi(x,z,t) = e \varphi_0 s(z) e^{i(qx-\omega t)}$, where $\varphi_0$ is the amplitude of potential on the top layer and $s(z)$ is the dimensionless 'shape function' (see Fig.~1 b).

A relatively simple form of the $\hat{\mathcal T}$-matrix allows us to treat the tunneling non-perturbatively~\cite{vasko_book}. The good quantum numbers of the eigen states of $\hat H_0$ are the in-plane electron momentum ${\bf p}$, the band index $s = \pm 1$ ($s=+1$ for the conduction and $s = -1$ for the valence band), and an extra index $l = \pm 1$ governing the $z$-localization of electron wave function. At large bias voltage, $l=+1$ and $l=-1$ correspond to the states localized on the top and bottom layers, respectively, while at small bias $l=+1$ and $l=-1$ describe the anti-symmetric and symmetric states of the tunnel-coupled quantum wells. The eigen energies of these states are
\begin{equation}
\varepsilon^{ls}_{\bf p} = s v_0 p + l \sqrt{\Omega^2 + \Delta^2/4}.
\end{equation}

The evaluation of tunnel current density is based on the relation $J_{\bot {\bf q}\omega} = g {\rm Tr} \left[J_{\bot \alpha \beta} \rho^{(1)}_{\beta\alpha}\delta_{{\bf p}+{\bf q},{\bf p}'}\right]$, where $\rho^{(1)}$ is the electron density matrix calculated up to the first order in electron-plasmon interaction, $g=4$ is the spin-valley degeneracy factor, and the indices $\alpha = \{{\bf p},l,s\}$, $\beta = \{{\bf p}',l',s'\}$ run over all quantum numbers. The explicit form of the tunnel current operator is found from the 'equation of motion', $\hat{J}_{\bot} = d \hat{Q}_t/dt = i [{\hat H},{\hat Q}_t]$, where ${\hat Q}_t$ is the operator of electron charge in the top layer. The density matrix ${\hat\rho}^{(1)}$ is found from the von Neumann equation
\begin{equation}
\label{Neumann}
i\frac{\partial \hat\rho^{(1)}}{\partial t} = [{\hat H}_0, \hat\rho^{(1)}] + [\hat{H}_{\rm{int}},\hat\rho^{(0)}],
\end{equation}
being interested in the linear response, we commute only the zeroth-order density matrix $\rho^{(0)}$ with $\hat{H}_{\rm{int}}$. Considering the harmonic time dependence, Eq.~(\ref{Neumann}) is readily solved in the diagonal basis of $\hat H_0$ leading to a closed-form expression for the frequency- dependent non-local tunnel conductance
\begin{equation}
\label{Tun_Current}
\displaystyle{ G_{ {\bf q},\omega} = i g e^2 \Omega \sum\limits_{
\begin{smallmatrix}
\bf {p},\bf {p}' = \bf {p} + \bf {q}\\
l\neq l', s, s'
\end{smallmatrix}
}
{ \frac{ u^{ss'}_{\bf pp'}  s_{ll'}  \left[ f^{sl}_{\bf p} - f^{s'l'}_{{\bf p}'} \right] }
        {\omega + i \delta - [\varepsilon^{sl}_{\bf p} - \varepsilon^{s'l'}_{{\bf p}'}] } 
 }.}
\end{equation}
Here, $s_{ll'}$ are the overlap integrals between the SP field profile and the $z$-components of the electron wave functions, $s_{ll'} = \int_{-\infty}^{\infty} dz \Psi^*_l(z) s(z) \Psi_{l'}(z)$, $u^{ss'}_{\bf pp'}$ are the overlap factors between the chiral envelope wave functions in graphene, $u^{ss'}_{\bf pp'} = [1 + ss' \cos \theta_{\bf pp'}]/2$, and $f^{sl}_{\bf p}$ are the occupation numbers of the eigen states of given by the Fermi functions shifted by the applied bias $eV$ in the energy scale. 

At this point, it is noteworthy to mention the similarity of Eq.~(\ref{Tun_Current}) and the expressions for the graphene polarizability \cite{Das_Sarma_Plasmons}. The latter diverge on the 'Dirac cone' as $[\omega^2 - q^2 v_0^2]^{-1/2}$, and a similar singularity appears in the tunnel conductance (\ref{Tun_Current}), except the frequency $\omega$ should be replaced with $\omega - \sqrt{\Delta^2 + 4\Omega^2} $. This singularity is simply elucidated after extracting the real part of Eq.~(\ref{Tun_Current}) and passing to the elliptic coordinates, which leads us to the following expression
\begin{equation}
\label{ReGtun_integrated}
{\rm{Re}} G_{\bot {\bf q},\omega} = - \frac{e^2}{2\pi \hbar}  \frac{q^2 s_\pm \Omega \mathcal{I}}{\sqrt{(qv_0)^2 - (e\tilde{V}_{12} - \omega)^2}} ,
\end{equation}
where we have introduced the effective energy shift between the Dirac points $e{\tilde V}_{12} = \sqrt{\Delta^2 + 4\Omega^2}$, and a dimensionless non-singular function $\mathcal{I}$ appearing as a result of the Fermi distribution integration,
\begin{equation}
\mathcal{I} = \sinh\left(\frac{eV -\omega}{T}\right)\int\limits_{1}^{\infty}{\frac{\sqrt{t^2-1}dt}{\cosh\left(\frac{qv_0}{T}t\right)+\cosh\left(\frac{eV - \omega}{T}\right)}}.
\end{equation}

The real part of the tunnel conductance is negative provided $\omega < e V$, which implies that the tunneling electrons rather loose their energy by emitting a plasmon than get the energy by plasmon absorption. At the same time, the tunnel transitions from top layer to the bottom one accompanied by SP absorption are prohibited by the energy conservation law as far as the SP velocity exceeds the Fermi velocity. Still, just like in most tunnel phenomena, the negative conductivity due to the tunneling is proportional to a small exponent $e^{-2\kappa d}$ ($\kappa$ is the decay constant of the electron wave function) that eventually enters $s_\pm$ and $\Omega$. To enhance the negative conductance, the materials with small effective (tunneling) mass $m^*$ and small band offset $U_b$ with respect to graphene have to be used. Boron nitride ($U_b = 1.5$ eV, $m^* = 0.5 m_0$) is not the best candidate for the realization of the net SP gain, while chalcogenides of molybdenum MoS$_2$  ($U_b = 0.29$ eV, $m^* = 0.43 m_0$) and tungsten WS$_2$ ($U_b = 0.4$ eV, $m^* = 0.28 m_0$)~\cite{Band_parameters_MOS2} demonstrate significantly stronger tunneling. 

The smallness of tunnel current is to a considerable extent compensated by a singularity in plasmon gain, which occurs provided $e\tilde{V}_{12} = \omega +q v_0 = \omega (1 + v_0/s)$. The structure of the ${\bf p}$-integral in Eq.~(\ref{Tun_Current}) shows that the singularity comes from the tunneling between electron states with collinear momenta in the neighboring layers. Once the carrier dispersion is linear, these states have the same velocity and thus interact for an infinitely long time~\cite{Fritz_PRB}. The carrier scattering naturally destroys this coherence making the singular tunnel conductance finite. To account for this effect quantitatively, we add the imaginary self-energy corrections ${\rm Im}\Sigma({\bf p},E)$ to the quasi-particle energies in (\ref{Tun_Current}). These corrections appear as a result of electron-phonon scattering, the latter also leading to the finite Drude absorption. For simplicity, we take ${\rm Im}\Sigma({\bf p},E)$ to be energy- and momentum independent and equal to its value at the Fermi-surface ${\rm Im}\Sigma({\bf p}_F,\varepsilon_F) = \nu$,
\begin{equation}
\label{Phonon_collision}
\nu = \frac{\varepsilon_F T D^2}{ 4 v_0^2 \rho c^2},
\end{equation}
where $D \simeq 30$ eV is the deformation potential of graphene~\cite{PRL_Bolotin}, $\rho  = 7.6 \times 10^{-7}$ kg/m$^2$ is its mass density, and $c \approx 0.02 v_0$ is the sound velocity~\cite{Vasko-Ryzhii}.


To judge on the possibility of the net plasmon gain, one has also to evaluate the frequency-dependent non-local in-plane conductivity, $\sigma_{{\bf q},\omega}$. In the absence of electron collisions, one can readily apply the Kubo linear response theory and obtain the well-known result including inter- and intraband contributions~\cite{Falkovsky-Varlamov}. The real part of interband optical conductivity ($q = 0$) is universal at high frequencies, ${\rm Re}\sigma_{\rm inter} = e^2/4\hbar$. However, at finite wave vector -- which is the case of plasmons -- the frequency dependence of conductivity becomes more complicated,
\begin{equation}
\label{Inter-simple}
{\rm{Re}}\sigma_{\rm{inter}}({\bf q},\omega) \approx \frac{e^2}{4\hbar}\frac{\omega}{\sqrt{\omega^2 - q^2 v_0^2}}\left[f_0\left(-\omega/2 \right) - f_0\left(\omega/2 \right) \right].
\end{equation}
The neglect of the spatial dispersion results in an underestimate of ${\rm Re}\sigma_{\rm inter}$ and, hence, plasmon damping. This is especially crucial for the acoustic modes which velocity just slightly exceeds the Fermi velocity and the respective interband conductivity is in a dangerous vicinity of the singularity at $\omega = q v_0$.

Considering the intraband (Drude) conductivity, a special care should be taken to account correctly both for the spatial dispersion and finite carrier relaxation time~\cite{Principi_plasmon_loss_hBN}. Apparently, the effects of carrier collisions cannot be included by a simple replacement $\omega \rightarrow \omega + i \nu$ in the Kubo-like expression for the collisionless conductivity~\cite{Mermin-Lindhard_dielectric_Function}. Even in the quasi-classical formalism of Boltzmann equation (which is justified in the case of interest, $q \ll q_F$, $\omega \ll \varepsilon_F$), the collision integral in the $\tau$-approximation violates the particle conservation. To avoid those difficulties, we have evalauted the conductivity by solving the kinetic equation with the particle-conserving Bhatnagar-Gross-Krook collision integral~\cite{BGK-collisions} in the right-hand side. The respective intraband dynamic conductivity reads
\begin{equation}
\label{Sigma_intra}
\displaystyle{\sigma_{\rm{intra}}({\bf q},\omega) = \frac{i e^2 \tilde\varepsilon_F}{\pi^2\omega}\displaystyle{\frac{J_2(\frac{qv_0}{\omega},\frac{\nu}{\omega})}{1 - \frac{i}{2\pi}\frac{\nu}{\omega}\frac{qv_0}{\omega}J_1(\frac{qv_0}{\omega},\frac{\nu}{\omega})}},}
\end{equation}
where $J_{\rm n}(a,b) = \int_0^{2\pi}{d\theta \cos^n\theta}/[1 - a \cos\theta + i b]$, $\tilde\varepsilon_F = T \ln(1+e^{\varepsilon_F/T})$, and $\nu$ is the collision frequency limited by the electron-phonon scattering~\cite{Koppens_nano_imaging_hBN} given essentially by Eq.~(\ref{Phonon_collision}). At zero wave vector, $J_1 = 0$, and we restore the Boltzmann conductivity $\sigma_{\rm{intra}}(0,\omega) = i e^2 \tilde\varepsilon_F/\pi(\omega + i \nu)$, while for nonzero $q$ the real pat of intraband conductivity typically appears to be larger than its zero-$q$ value. This difference between no-local and local conductivities agrees with the experimentally observed distinction between the plasmon lifetime and the Boltzmann relaxation time $\nu^{-1}$~\cite{Koppens_nano_imaging_hBN}.

\begin{figure}[ht]
\includegraphics[width=0.9\linewidth]{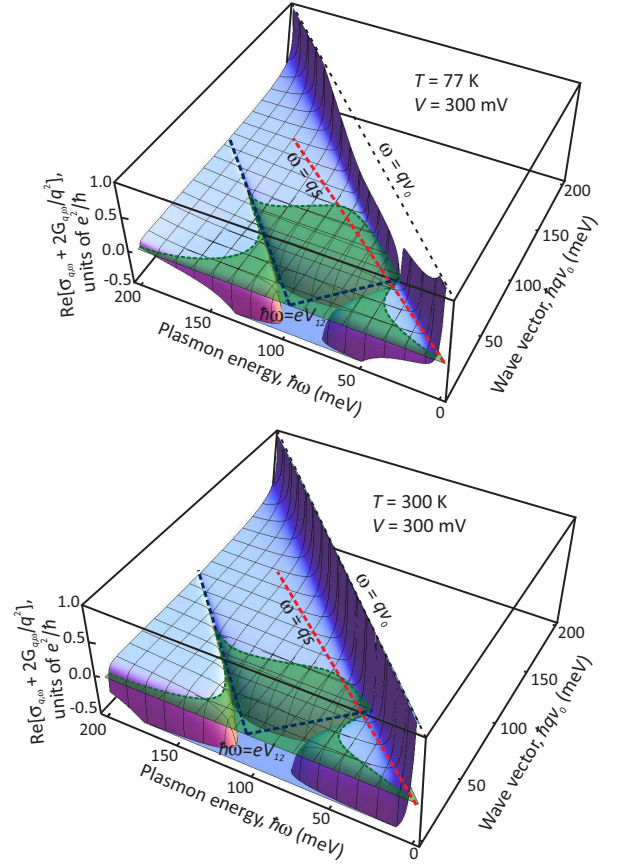}
\caption{\label{NetSigmaG} Net ''effective conductivity'' as it appears in the SP dispersion law, ${\rm Re}[\sigma_{q,\omega} + 2 G_{q,\omega}/q^2]$ (in the units of $e^2/\hbar$) vs. frequency and wave vector at two different temperatures: $T = 77$ K (top) and $T = 300$ K (bottom). The barrier layer is 2.5 nm WS$_2$. The singularities of the interlayer tunnel conductance, $\hbar\omega = e V_{12}\pm q v_0$ are shown with blue lines, the dispersion of acoustic SP is shown with red line. In the region filled with green, the net effective conductivity is negative. 
}
\end{figure}

\begin{figure}[ht]
\includegraphics[width=0.9\linewidth]{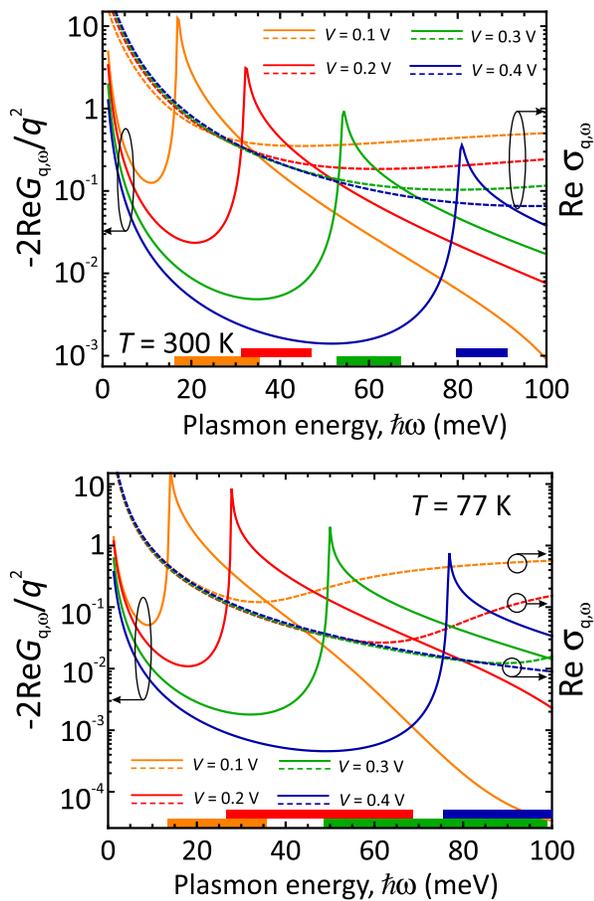}
\caption{\label{CW_gain_loss} Comparison of the real part of in-plane graphene conductivity ${\rm Re}[\sigma_{\rm inter} + \sigma_{\rm intra}]$ governing the plasmon loss (normalized to $e^2/\hbar$, dashed lines), and the real part of the effective tunnel conductance $-2 {\rm Re}G/q^2$ governing the plasmon gain (solid lines). The barrier layer is 2.5 nm WS$_2$. In the frequency ranges where ${\rm Re}[\sigma_{\rm inter} + \sigma_{\rm intra} - 2 G /q^2] < 0$, the plasmons are amplified instead of being damped. These ranges are marked with color bars near the frequency axis.
}
\end{figure}

With these prerequisites, we are able to compare the plasmon gain due to the tunneling and loss due to the Drude and interband absorption. Namely, if in some frequency range the 'effective conductivity'  $\sigma_{\rm eff} = {\rm Re}[\sigma_{\rm inter}+\sigma_{\rm intra} - 2 G/q^2] < 0$, then the plasmon gain exceeds loss. The self-excitation of plasmons with frequencies satisfying the net gain condition is possible. Figure \ref{NetSigmaG} shows the wave vector- and frequency dependence of the effective conductivity at the room and nitrogen temperatures with the region under the green plane corresponding to the negative values. The two sets of singularities are clearly present in the conductivity spectra: the absorption singularity at $\omega = q v_0$ comes from the in-plane conductivity, and the gain singularity at $\omega = e\tilde V_{12} \pm q v_0$ comes from the enhanced tunneling between the collinear states. As the wave vector approaches zero, the effective conductance grows indefinitely due to the presence of $q^2$ in the denominator, this will not, however, lead to the infinite plasmon gain as the acoustic SP spectra develop a gap $\Delta \omega \simeq \Omega$ at small wave vectors~\cite{DasSarma-PRL-tunnel-plasmon}. On the other hand, there is a wide range of frequencies $\omega \gtrsim \Omega$ where the dispersion of acoustic SPs in not strongly affected by tunneling, but the real part of net effective conductivity is negative, which justifies the possibility of gain.

The spectra of in-plane conductivity ${\rm Re}[\sigma_{\rm inter}+\sigma_{\rm intra}$ and effective tunnel conductance $- 2 {\rm Re}[G/q^2]$ are compared in Fig.~\ref{CW_gain_loss} for the wave vectors satisfying the acoustic plasmon dispersion $\omega = s q$. At room temperature (upper panel) the self-excitation is possible in a narrow ($\sim 15-20$ meV) vicinity of the tunneling resonance frequency, while at higher frequencies the strong interband absorption surpasses the gain. At lower temperature of 77 K (lower panel) the range of frequencies corresponding to the net SP gain is significantly broader, while the peak tunnel conductance is higher. The broadening of the gain region is due to efficient low-temperature Pauli blocking of the interband transitions with energy below the double Fermi energy. The sharpening of the resonant peak is attributed to the phonon freezing-out at low temperatures, which increases the quasiparticle lifetime and resonant gain. It is also noteworthy that the peak in the negative tunnel conductance occurring at $\omega_{\rm res} = e {\tilde V}_{12}/(1 + v_0/s) \approx e {\tilde V}_{12}/2$ falls into the 'transparency window' of graphene where the Drude absorption is low ($\omega_{\rm res} \gg \nu$) but the interband absorption is still blocked ($\omega_{\rm res} < 2 \varepsilon_F$).

The efficiency of SP excitation can be further improved in the gated double-graphene layer structures, where the energy shift between Dirac points $\Delta$ and the carrier density can be controlled independently. At the fixed carrier density, the plasmon gain increases with reduction in $\Delta$ as the energies of the $l = +1$ and $l = -1$ states get closer in energy scale resulting in enhanced tunnel coupling.

In realistic graphene-based tunnel structures, the adjacent layers are always slightly misaligned in real space, which results in the misalignment of Dirac cones in the $k$-space. The finiteness of SP wave vector can, to some extent, compensate this misalignment. If the Dirac cones in adjacent layers are separated by a wave vector ${\bf q}_M$, the tunneling resonance will be retained, though the position of the resonant peak will depend on the direction of plasmon propagation $\bf {q}$. The resonant condition in this case is easily shown to be
\begin{equation}
|{\bf q} + {\bf q}_M|^2 v_0^2 = (eV_{12} - \omega)^2,
\end{equation}
and the expressions for the real part of tunnel conductance are obtained from Eq.~(\ref{ReGtun_integrated}) by a simple replacement $q \rightarrow |{\bf q} + {\bf q}_M|$. The case of the plasmon-assisted resonant tunneling is thus radically different from the photon-assisted tunneling~\cite{Ryzhii_DGL_laser}, where even a slight misalignment breaks the resonance due to the negligible photon momentum. The robustness of SP emission against slight misalignment correlates with the broadness of gain (and emission) spectra extending above the resonant frequency. At the same time, the spectra of phonons emitted upon tunneling are predicted to be Lorentzian their width being limited by the carrier relaxation rate. For these reason, the inelastic tunneling accompanied by the emission of surface plasmons rather than photons may largely contribute to the spontaneous terahertz emission from the double graphene layer structures observed in~\cite{THz_emission_DGL_experim}. The mechanism of electromagnetic radiation in case of plasmon excitation could be either the dipole radiation of the whole heterostructure, or the radiative decay of plasmons scattered by the edges.

In conclusion, we have theoretically studied the frequency-dependent non-local tunnel conductivity in the biased graphene layers. In a wide range of frequencies, the real part of the tunnel conductivity is negative due to the dominance of inelastic tunneling with surface plasmon emission over the absorption.  Moreover, at frequencies satisfying the condition $\hbar \omega = e V_{12} \pm q v_0$, where $V_{12}$ is the interlayer potential difference and $q$ is the plasmon wave vector, the absolute value of negative tunnel conductivity is resonantly large, which is a manifestation of enhanced tunneling between electron states with collinear momenta in the neighboring layers. A detailed comparison of the surface plasmon loss due to the interband and Drude absorption and gain due to the inelastic tunneling shows the possibility of the net gain in sufficiently thin tunnel structures.

The work of DS was supported by the grant \# 14-07-31315 of the Russian Foundation of Basic Research. The work at RIEC was supported by the Japan Society for Promotion of Science (Grant-in-Aid for Specially Promoted Research No. 23000008).

\section{Appendix}
\appendix
\setcounter{equation}{0}
\renewcommand{\theequation} {A\arabic{equation}}

\setcounter{figure}{0}
\renewcommand{\thefigure} {A\arabic{figure}}

\section{Evaluation of the in-plane conductivity}
The expression for the interband part of conductivity is readily obtained from the Kubo theory~\cite{Falkovsky-Varlamov}
\begin{multline}
\sigma_{\rm{inter}}({\bf q},\omega) = \\
\frac{2i e^2 \omega }{\pi^2} 
\sum_{{\bf p},s\neq s'}
{
    \frac{\left| {\bf v}^{\rm{ss'}}_{ {\bf p}{\bf p}'}\right|^2 
    \left[f^{s'}_{{\bf p}'} - f^{s}_{{\bf p}}\right] }
    { \left[ \epsilon^{s'}_{{\bf p}'} - \epsilon^{s}_{{\bf p}} \right] 
      \left [ \omega + i \delta - (  \epsilon^{s'}_{{\bf p}} - \epsilon^{s}_{{\bf p}} ) \right] }}.
\end{multline}
Here ${\bf v}^{\rm{ss'}}_{ {\bf p}+,{\bf p}-}$ is the matrix element of velocity operator in graphene $\hat{\bf v} = v_0 {\bf \sigma}$, ${\bf p}_{\pm} = {\bf p} \pm {\bf q}/2$, $f^{s}_{\bf p}$ is the electron distribution functions in the valence ($s=-1$) and the conduction ($s=+1$) band, and $\epsilon^{s}_{\bf p}$ is the dispersion law in the $s$-th band. Known the eigen functions of graphene Hamiltonian $\hat{H}_G = v_0 {\bf \sigma p}$,
\begin{equation}
\ket{ s {\bf p}} =\frac{1}{\sqrt{2}}\left( 
\begin{aligned}
  & {{e}^{-i{{\theta }_{\bf{p}}}/2}} \\ 
 & {s{e}^{i{{\theta }_{\bf{p}}}/2}} \\ 
\end{aligned} \right) e^{i {\bf p r}},
\end{equation}
one readily finds $\bra{c {\bf p}_-} \hat{v}_x \ket{v {\bf p}_+} = i v_0 \sin\left[(\theta_{{\bf p}+} + \theta_{{\bf p}-})/2\right]$. The subsequent calculations are conveniently performed in the elliptic coordinates
\begin{equation}
{\bf p} =\frac{q}{2}\left\{ \cosh u\cos v, \sinh u \sin v \right\}.
\end{equation}
In these coordinates $|{\bf p}_\pm| = (q/2)[\cosh u \pm \cos v]$, $|\bra{c {\bf p}_-} \hat{v}_x \ket{v {\bf p}_+}|^2 dp_x dp_y = (q v_0/2)^2 \cosh^2u\sin^2v du dv$. The real part of conductivity is readily extracted with Sokhotski theorem, which leads us to
\begin{multline}
\label{Inter-exact}
{\rm{Re}}\sigma_{\rm{inter}}({\bf q},\omega) = \frac{e^2}{2\pi}\frac{\omega}{\sqrt{\omega^2 - q^2 v_0^2}}\int\limits_0^\pi dv \sin^2v \times \\
\left\{f_0\left[-\frac{\omega}{2} + \frac{q v_0}{2} \cos v \right] - f_0\left[\frac{\omega}{2} + \frac{q v_0}{2} \cos v \right] \right\}.
\end{multline}

\begin{figure}[ht]
\includegraphics[width=0.9\linewidth]{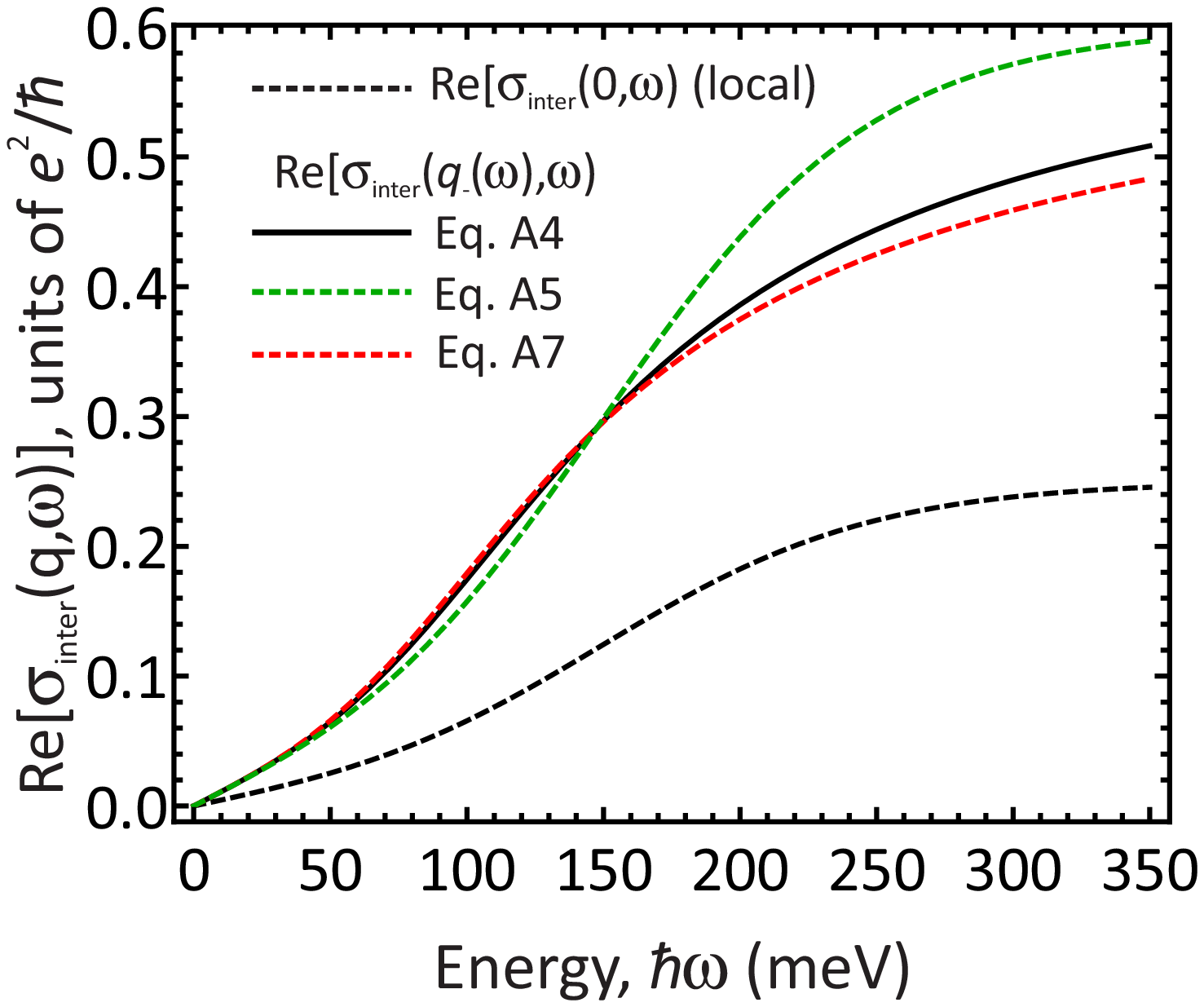}
\caption{\label{Re_Sigma} Real part of the interband in-plane conductivity vs. frequency calculated using exact expression (\ref{Inter-exact}) and its various approximations: local ($q=0$), black dashed line; Eq.~(\ref{Inter-approx1}), green dashed line; Eq.~(\ref{Inter-approx2}), red dashed line. For the non-local results, the wave vector $q$ satisfies the dispersion of acoustic plasmon $q_-(\omega) = \omega/s$. Temperature $T=300$ K, Fermi energy $\varepsilon_F = 75$ meV, $s = 1.1 v_0$.
}
\end{figure}

To proceed further, we note that in the domain of interest $\omega > q v_0$ one always has $ q v_0 \cos v < w$. As a simplest approximation, one can set $q = 0$ in the arguments of distribution function and obtain
\begin{multline}
\label{Inter-approx1}
{\rm{Re}}\sigma_{\rm{inter}}({\bf q},\omega) \approx \frac{e^2}{4\hbar}\frac{\omega}{\sqrt{\omega^2 - q^2 v_0^2}}\left[f_0\left(-\omega/2 \right) - f_0\left(\omega/2 \right) \right].
\end{multline}
The agreement between approximate analytical expression (\ref{Inter-approx1}) and exact integral relation (\ref{Inter-exact}) is fair as far as $qv_0$ is not too close to $\omega$. A better agreement can be obtained after the change of variable $\cos u = t$ in Eq.~(\ref{Inter-exact}) by noting that $\sin v \equiv  \sqrt{1-t^2}$ varies rapidly over the integration domain $t \in [-1;1]$, while the remainder $g(t) = f_0[-(\omega - qt)/2] - f_0[(\omega + qt)/2]$ varies slowly. Thus, one can make the following approximation
\begin{equation}
\int\limits_{-1}^{1}{g(t)\sqrt{1-t^2} dt} \approx \frac{1}{2} \int\limits_{-1}^{1}{ \sqrt{1-t^2} dt}\int\limits_{-1}^{1}{g(t) dt}.
\end{equation}
The following approximation for the conductivity holds
\begin{multline}
\label{Inter-approx2}
{\rm{Re}}\sigma_{\rm{inter}}({\bf q},\omega) \approx \frac{e^2}{4\hbar} \frac{T}{q v_0}\frac{\omega}{\sqrt{\omega^2 - q^2 v_0^2}}\times\\
\ln\frac{\cosh \frac{\varepsilon_F}{T} + \cosh \frac{\omega + qv_0}{2T}}{\cosh \frac{\varepsilon_F}{T} + \cosh \frac{\omega - qv_0}{2T}}.
\end{multline}

Figure \ref{Re_Sigma} shows the calculated interband conductivities using the exact (\ref{Inter-exact}) and approximate expressions (\ref{Inter-approx1}) and (\ref{Inter-approx2}) as well as the local ($q = 0$) approximation. Clearly, the neglect of spatial dispersion in the case of acoustic SPs with velocity slightly exceeding the Fermi velocity results in an underestimation of the damping.

We now pass to the in-plane conductivity associated with the intraband transitions, ${\rm{Re}}\sigma_{\rm{intra}}({\bf q},\omega)$. We shall restrict ourselves to the classical description of the intraband electron motion which is justified at frequencies $\hbar\omega \ll \varepsilon_F$, $q \ll q_F$. The effect of plasmon tunneling gain occurs in this frequency domain, otherwise, the interband damping of SPs takes place [see Eq.~\ref{Inter-simple}]. To account for the carrier relaxation in the non-local ($q \neq 0$) case we adopt the formalism of kinetic equation with the particle-conserving Bhatnagar-Gross-Krook collision integral ~\cite{BGK-collisions} in the right-hand side,
\begin{multline}
\label{Kinetic-BGK}
-i\omega \delta f_{\bf q} ({\bf p}) +i{\bf q v} \delta f_{\bf q} ( {\bf p} ) + i e {\bf q v}{\varphi_{q}}\frac{\partial {f_0}}{\partial \varepsilon }=\\
-\nu \left[ \delta {f_{\bf q}} ( {\bf p} ) + \frac{d{{\varepsilon }_{F}}}{dn}\frac{\partial {f_0}}{\partial \varepsilon }\delta {n_{\bf q}} \right].
\end{multline}
Here $\delta f_{\bf q} (\bf{p})$ is the sought-for field-dependent correction to the equilibrium electron distribution function $f_0$, $\delta n_{\bf q}$ is the respective correction to the electron density, ${\bf v} = v_0 {\bf p}/p$ is the quasi-particle velocity, and $\nu$ is the electron collision frequency which is assumed to be energy-independent. The current density, associated with the distribution function $\delta f_{\bf q} (\bf{p})$ reads:
\begin{equation}
\label{Current_BGK}
\delta {\bf j}_{\bf q} = - e g \sum_{\bf p}{{\bf v} \frac{df_0}{d\varepsilon} \frac{i e {\bf qv} \varphi_{\bf q} - i \nu \frac{d\varepsilon_F}{dn}\delta n_{\bf q} }{\omega + i \nu - {\bf q v}} }.
\end{equation}
Recalling the relation between small-signal variations of density and current, $\omega \delta n_{\bf q} = q \delta {\bf j}_{\bf q}$, and evaluating the integrals in Eq.~(\ref{Current_BGK}), we find the conductivity given by Eq.~(\ref{Sigma_intra}).

\section{Spectra of plasmons coupled to the double graphene layer heterostructures}

The plasmon spectra are obtained by a self-consistent solution of the Poisson's equation 
\begin{multline}
\label{Poisson}
-q^2 \delta \varphi_{\bf q} (z) + \frac{\partial^2 \delta \varphi_{\bf q} (z)}{\partial z^2} = \\
-\frac{4\pi}{\kappa}\left[ \delta \rho_{t,{\bf q}}\delta(z-d/2) + \delta \rho_{b,{\bf q}}\delta(z+d/2) \right],
\end{multline}
the continuity equations
\begin{equation}
-i\omega\delta \rho_{\bf q} + {\bf q}{\delta {\bf j}_{\bf q}}=0,
\end{equation}
and the linear-response relation between current density and electric field, $\delta {\bf j}_{\bf q} = \sigma_{{\bf q},\omega} {\bf E}_{\bf q}$. Here ${\bf q}$ is the two-dimensional plasmon wave vector, $d$ is the distance between layers, $\kappa$ is the background dielectric permittivity, $\delta \rho_{t,{\bf q}}$ and $\delta \rho_{b,{\bf q}}$ are the small-signal variations of charge density in the top and bottom layers, respectively. In the absence of built-in voltage, due to the electron-hole symmetry, the charge densities in the layers are equal in modulus an opposite in sign, moreover, the layer conductivities are equal. This allows us to seek for the solutions of Eq.~(\ref{Poisson}) being symmetric and anti-symmetric with respect to the electric potential. A straightforward calculation brings us to the following dispersions~\cite{Voltage_controlled}
\begin{equation}
\label{AntiSymmetric-disp}
-\frac{i\omega \kappa }{4\pi }\left( 1+\coth \frac{qd}{2} \right)+q{{\sigma }_{{\bf q}\omega }}=0
\end{equation}
for the antisymmetric (acoustic) mode, and
\begin{equation}
\label{Symmetric-disp}
-\frac{i\omega \kappa }{4\pi }\left( 1+\tanh \frac{qd}{2} \right)+q{{\sigma }_{{\bf q}\omega }}=0
\end{equation}
for the symmetric (optical mode). Being interested in the long-wavelength limit, $q d/2 \ll 1$, we perform the expansions $1+\coth qd/2 \approx 2/qd$, $1+\tanh qd/2 \approx 1$. In the same limit, the conductivity is essentially classical, moreover, the interband transitions do not affect the low-energy part of the spectra. With these assumptions, we use the following (collisionless) approximation for the conductivity
\begin{equation}
{{\sigma }_{\bf{q}\omega }}=i g \frac{{e^2}}{\hbar }\frac{{\tilde{\varepsilon}}_F}{2\pi }\frac{\omega }{q^2v_0^2}\left[ \frac{\omega }{\sqrt{{\omega^2}- q^2 v_0^2}}-1 \right],
\end{equation}
which follows readily from Eq.~(\ref{Sigma_intra}) after setting $\nu = 0$. Equation~(\ref{AntiSymmetric-disp}) admits an analytical solution, which is a sound-like dispersion
\begin{equation}
\omega_- = v_0 \frac{1 + 4 \alpha_c q_F d}{\sqrt{1 + 8 \alpha_c q_F d}} q.
\end{equation}
Here, we have introduced the Fermi wave vector $q_F = \tilde\varepsilon_F/v_0$. The velocity of the acoustic mode always exceeds the Fermi velocity thus preventing the Landau damping. The dispersion equation for the optical mode $\omega_+(q)$ is cubic, however, in the long-wavelength limit the spatial dispersion of conductivity can be neglected as the phase velocity of this mode significantly exceeds the Fermi velocity. The approximate relation for $\omega_+(q)$ has the following form
\begin{equation}
\omega_+ \approx v_0 \sqrt{4 \alpha_c q q_F}.
\end{equation}

\begin{figure}[ht]
\includegraphics[width=0.9\linewidth]{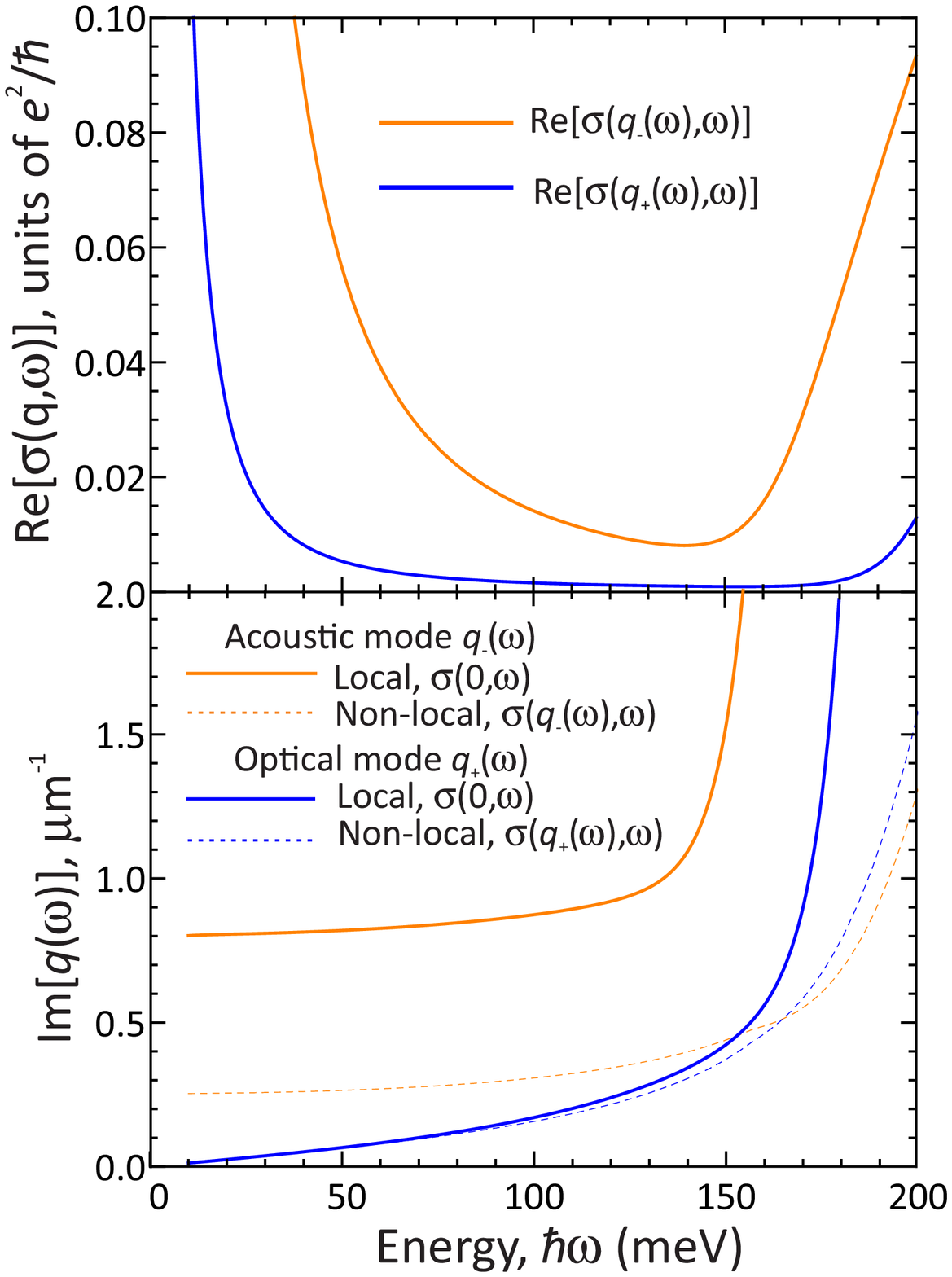}
\caption{\label{Net_Sigma} Top panel: real part of the non-local net in-plane conductivity (Drude + interband) vs. frequency calculated for the wave vectors satisfying the dispersion of the acoustic mode $q_(\omega)$ (orange line) and the optical mode $q_+(\omega)$. Temperature $T=77$ K, Fermi energy $\varepsilon_F = 150$ meV, $s = 1.2 v_0$. Bottom panel: calculated imaginary  parts of the propagation constant for the acoustic and optical modes calculated using the local and non-local expressions for the conductivity
}
\end{figure}

To account for the damping of the propagating plasmon one can express the wave vector $q$ from Eqs.~(\ref{AntiSymmetric-disp}) and (\ref{Symmetric-disp}) and consider the real part of conductivity along with the imaginary one. The results of such calculations are shown in Fig.~\ref{Net_Sigma} along with the frequency dependencies of the net conductivity. The acoustic mode is typically damped more heavily than optical one the two effects being responsible for that fact. First of all, the velocity of optical mode is higher than that of acoustic one, hence, at a fixed scattering time, the free path of optical mode is higher. The second reason is the increase in the real part of conductivity at high spatial dispersion discussed above. The use of 'local' expressions for conductivity results in a factor of two error in the free path of acoustic mode, which is also illustrated in Fig.~(~\ref{Net_Sigma}).

In the following calculations we shall also require the spatial dependence of the plasmon potential in the acoustic mode, which can be obtained from (\ref{Poisson}). It is convenient to present it as 
\begin{equation}
\label{Plasm-potential}
\delta\varphi_{\bf q}(z) = \delta\varphi_0 s(z),
\end{equation}
where $\varphi_0$ is the electric potential on the top layer. The shape function has the following form
\begin{equation}
\label{Shape-function}
s\left( z \right)=\left\{ \begin{aligned}
  & {{e}^{-k\left( z+d/2 \right)}},\,\,z<-d/2, \\ 
 & -\frac{\sinh \left( qz/2 \right)}{\sinh \left( qd/2 \right)},\,\,\left| z \right|<d/2, \\ 
 & -{{e}^{-k\left( z-d/2 \right)}},\,\,z>d/2. \\ 
\end{aligned} \right.
\end{equation}

\section{Estimate of the effective tight-binding parameters}
The tight-binding Hamiltonian of the tunnel-coupled graphene layers in the absence of the propagating plasmon [${\hat H}_0$ in Eq.~(\ref{Hamiltonian})] constitutes the blocks describing isolated graphene layers ${\hat H}_{G\pm}$ and the block describing tunnel hopping $\hat{\mathcal T}$. This Hamiltonian acts on a four-component wave function 
\begin{equation}
\label{Psi_TB}
\vec\psi = \left\{\psi_{A,t},\psi_{B,t},\psi_{A,b},\psi_{B,b}\right\}^{\rm T}
\end{equation}
which components are the probability amplitudes of finding an electron on a definite layer and on the definite lattice cites, $A$ or $B$. Such description of electron states is common for graphene bilayer; moreover, the elements of tunneling matrix have been evaluated for different twist angles between single layers constituting a bilayer. In more comprehensive theories, the $\hat{\mathcal T}$-matrix is affected by the band structure of dielectric layer. Here, for the sake of analytical traceability, we choose the tunneling matrix in its simplest form which is applicable to the AA-stacked perfectly aligned graphene bilayer, $\hat{\mathcal T} = \Omega \hat I$, where $\Omega$ can be interpreted as the tunnel hopping frequency. 

To estimate its value, we switch for a while from the tight binding to the continuum description of electron states in the $z$-direction. We model each graphene layer with a delta-well $U(z) = A \delta (z - z_{t,b})$, where $A$ is the potential strength chosen to provide a correct value of electron work function $U_b$ from graphene to the background dielectric,
\begin{equation}
A = 2\sqrt{\frac{\hbar^2 U_b}{2m^*}},
\end{equation}
where $m^*$ is the effective electron mass in the dielectric. The effective Schrodinger equation in the presence of voltage bias $\Delta/e$ between graphene layers takes on the following form
\begin{multline}
-\frac{\hbar^2}{2m^*}\frac{\partial^2\Psi(z)}{\partial z^2} + 2\sqrt{\frac{\hbar^2 U_b}{2m^*}}\left[\delta(z-d/2) + \delta(z+d/2)\right]\Psi(z) +\\
+U_F(z)\Psi(z) = E \Psi(z),
\end{multline}
where $U_F$ is the potential energy created by the applied field
\begin{equation}
U_F\left( z \right)=\frac{\Delta}{2} \left\{ \begin{aligned}
  & 1,\,\,z<-d/2, \\ 
 & 2 z/d,\,\,\left| z \right|<d/2, \\ 
 & -1,\,\,z>d/2. \\ 
\end{aligned} \right.
\end{equation}
The solutions of effective Schrodinger equation represent decaying exponents at $|z| > d/2$, and a linear combination of Airy functions in the middle region $|z| < d/2$
\begin{equation}
\Psi_M(z) = C {\rm Ai} \left(-z/a + \varepsilon \right) + D {\rm Bi} \left(-z/a + \varepsilon \right),
\end{equation}
where $\varepsilon= 2m^* |E| a^2/\hbar^2 $ is the dimensionless energy and $a = (\hbar^2 d/ 2 m^*\Delta)^{1/3}$ is the effecive length in the electric field. A straightforward matching of the wave functions at the graphene layers yields the dispersion equation
\begin{widetext}
\begin{equation}
\label{Energy_spectrum}
\det \left( \begin{matrix}
   {e^{-k_1 d/2}} & -\text{Ai}\left( d/2a+\varepsilon  \right) & -\text{Bi}\left( d/2a+\varepsilon  \right) & 0  \\
   \left(2 k_b - k_1 \right){{e}^{-{k_1}d/2}} & -\frac{1}{a}\text{Ai}'\left( d/2a+\varepsilon  \right) & -\frac{1}{a}\text{Bi}'\left( d/2a+\varepsilon  \right) & 0  \\
   0 & -\text{Ai}\left( -d/2a+\varepsilon  \right) & -\text{Bi}\left( -d/2a+\varepsilon  \right) & {{e}^{-{{k}_{2}}d/2}}  \\
   0 & -\frac{1}{a}\text{Ai}'\left( -d/2a+\varepsilon  \right) & -\frac{1}{a}\text{Bi}'\left( -d/2a+\varepsilon  \right) & \left( 2k_b - k_2 \right){e^{-k_2 d/2}}  \\
\end{matrix} \right)=0,
\end{equation}
\end{widetext}
where $k_b = \sqrt{2m^* U_b/\hbar^2}$ is the decay constant of the bound state wave function in a single delta-well, $k_1 =\sqrt{2m^* (E + \Delta/2)/\hbar^2}$, $k_2 =\sqrt{2m^* (E - \Delta/2)/\hbar^2}$ Equation (\ref{Energy_spectrum}) yields two energy levels $E_l$ ($l = \pm 1$) which cannot be expressed in a closed form via the parameters of the well. However, the dependence of $E_l$ on the energy separation between layers $\Delta$ is accurately described by
\begin{equation}
\label{Energy_spectrum_typical}
E_l (\Delta) = -U_b + \frac{l}{2} \sqrt{\left(E_{+1,\Delta=0} - E_{-1,\Delta=0}\right)^2 + \Delta^2}.
\end{equation}
The energy spectrum (\ref{Energy_spectrum_typical}) is typical for the tunnel coupled quantum wells~\cite{vasko_book}; the same functional dependence of energy levels on $\Delta$ is naturally obtained by diagonalizing the block Hamiltonian (\ref{Hamiltonian}),
\begin{equation}
\label{Energy_spectrum_block}
E_l (\Delta) = -U_b + l \sqrt{\Omega^2+ \frac{\Delta^2}{4}}.
\end{equation}
This allows us to estimate the tunnel coupling $\Omega$ as half the energy splitting of states in double graphene layer well in the absence of applied bias
\begin{equation}
\Omega=\frac{1}{2}\left[E_{+1,\Delta=0} - E_{-1,\Delta=0} \right].
\end{equation}

The $l$-index governs the $z$-localization of electron in a biased double quantum well. At large bias $\Delta \gg \Omega$, the delta-wells interact weakly, thus $l = +1$ corresponds to the state localized almost completely in the top layer and $l = -1$ corresponds to the electron in the bottom layer. At small bias $\Delta \approx \Omega$ the state $l = +1$ is anitisymmetric and $l = -1$ is symmetric.

\section{Electron-plasmon interaction and solution of the von Neumann equation}
The presence of plasmon propagating along the double graphene layer results in an additional potential energy of electron
\begin{equation}
\label{Plasm_energy}
H_{\rm int}({\bf r},t) = e \varphi(z) e^{i(q x-\omega t)},
\end{equation}
where we assume the direction of plasmon propagation to be along the $x$-axis, and the dependence of potential on the $z$-coordinate is given by Eqs.~(\ref{Plasm-potential}) and (\ref{Shape-function}). The additional terms in Hamiltonian due to the vector-potential are negligible as far as the speed of light substantially exceeds the plasmon velocity.

With our choice of the tight-binding basis functions (\ref{Psi_TB}) as those localized on a definite layer and on a definite lattice cite, we shall require 16 matrix elements of the potential energy (\ref{Plasm_energy}) connecting those basis states. However, it is more convenient to work out the matrix elements of (\ref{Plasm_energy}) connecting the {\it eigen} states of Hamiltonian (\ref{Hamiltonian}). The good quantum numbers of these states are the in-plane momentum ${\bf p}$, the band index $s = \pm 1$  ($+1$ for the conduction band and $-1$ for the valence band) and the $l$ - index discussed above. The respective matrix elements are
\begin{multline}
\bra{{\bf p},s,l}\hat H_{\rm int} \ket{{\bf p}'s'l'} = \\
\delta_{{\bf p},{\bf p}'-{\bf q}} u^{ss'}_{{\bf pp}'} e \varphi_0 \int_{-\infty}^{\infty}{\Psi^*_l(z) s(z) \Psi_{l'}(z)}.
\end{multline}

Having obtained the matrix elements of electron-plasmon interaction, we pass to the solution of the von Neumann equation~(\ref{Neumann}). Being interested in the linear response of electron system to the plasmon field, we decompose the density matrix as $\hat\rho = \hat\rho^{(0)} + \hat\rho^{(1)}$, where $\hat\rho^{(1)}$ is linear in the electron-plasmon interaction $\hat H_{\rm int}$. The calculations are conveniently performed in the basis of the eigenstates of $\hat H_0$ including the effects of tunneling non-perturbatively. In this basis, $[{\hat H}_0,{\hat \rho}^{(1)}]_{\alpha \beta} = (\varepsilon_\alpha - \varepsilon_\beta){\rho}^{(1)}_{\alpha \beta}$, and thus one readily writes down the solution for the density matrix
\begin{equation}
\rho^{(1)}_{\alpha\beta} = \frac{\left[\hat H_{\rm int} , \hat\rho^{(0)} \right]_{\alpha\beta}}{\omega + i\delta - (\varepsilon_\alpha - \varepsilon_\beta) }.
\end{equation}
The first-order correction is now expressed through the density matrix in the absence of plasmon field $\hat\rho^{(0)}$. A particular choice of $\hat\rho^{(0)}$ requires the solution of kinetic equation in the voltage-biased tunnel-coupled layers, however, in several limiting cases the situation is greatly simplified~\cite{Kazarinov_Suris}. If the tunneling rate $\Omega$ is slower than the electron energy relaxation rate $\nu_{\varepsilon}$ (e.g., due to phonons and carrier-carrier scattering), the quasi-equilibrium distribution function is established in each individual layer. In this situation, $\hat\rho^{(0)}$ is diagonal in the basis formed by the wave functions localized on top and bottom layers its elements being the respective Fermi distribution functions. In the other limiting case, when tunneling is stronger than scattering ($\Omega \gg \nu_\varepsilon$), the electron is 'collectivized' by the two layers, and the density matrix $\hat\rho^{(0)}$ is diagonal in the basis of ${\hat H}_0$-eigenstates. For the parameters used in our calculations, $\hbar\Omega\approx 10$ meV exceeds the relaxation rate $\hbar \nu\approx 1$ meV, and the latter limiting case is justified. Setting $\rho^{(0)}_{\alpha\beta} = f_\alpha \delta_{\alpha\beta}$, where $f$ is the Fermi distribution function, we find
\begin{multline}
\bra{{\bf p},s,l}{\hat \rho}^{(1)}\ket{{\bf p}'s'l'} = \\
= s_{ll'} u^{ss'}_{{\bf pp}'} \frac{f^{s'l'}_{{\bf p}'} - f^{sl}_{\bf p}}{\omega + i\delta - (\varepsilon^{sl}_{\bf p} - \varepsilon^{s'l'}_{{\bf p}'}) }.
\end{multline}

\bibliography{Bibliography}

\begin{thebibliography}{47}%
\makeatletter
\providecommand \@ifxundefined [1]{%
 \@ifx{#1\undefined}
}%
\providecommand \@ifnum [1]{%
 \ifnum #1\expandafter \@firstoftwo
 \else \expandafter \@secondoftwo
 \fi
}%
\providecommand \@ifx [1]{%
 \ifx #1\expandafter \@firstoftwo
 \else \expandafter \@secondoftwo
 \fi
}%
\providecommand \natexlab [1]{#1}%
\providecommand \enquote  [1]{``#1''}%
\providecommand \bibnamefont  [1]{#1}%
\providecommand \bibfnamefont [1]{#1}%
\providecommand \citenamefont [1]{#1}%
\providecommand \href@noop [0]{\@secondoftwo}%
\providecommand \href [0]{\begingroup \@sanitize@url \@href}%
\providecommand \@href[1]{\@@startlink{#1}\@@href}%
\providecommand \@@href[1]{\endgroup#1\@@endlink}%
\providecommand \@sanitize@url [0]{\catcode `\\12\catcode `\$12\catcode
  `\&12\catcode `\#12\catcode `\^12\catcode `\_12\catcode `\%12\relax}%
\providecommand \@@startlink[1]{}%
\providecommand \@@endlink[0]{}%
\providecommand \url  [0]{\begingroup\@sanitize@url \@url }%
\providecommand \@url [1]{\endgroup\@href {#1}{\urlprefix }}%
\providecommand \urlprefix  [0]{URL }%
\providecommand \Eprint [0]{\href }%
\providecommand \doibase [0]{http://dx.doi.org/}%
\providecommand \selectlanguage [0]{\@gobble}%
\providecommand \bibinfo  [0]{\@secondoftwo}%
\providecommand \bibfield  [0]{\@secondoftwo}%
\providecommand \translation [1]{[#1]}%
\providecommand \BibitemOpen [0]{}%
\providecommand \bibitemStop [0]{}%
\providecommand \bibitemNoStop [0]{.\EOS\space}%
\providecommand \EOS [0]{\spacefactor3000\relax}%
\providecommand \BibitemShut  [1]{\csname bibitem#1\endcsname}%
\let\auto@bib@innerbib\@empty
\bibitem [{\citenamefont {Grigorenko}\ \emph {et~al.}(2012)\citenamefont
  {Grigorenko}, \citenamefont {Polini},\ and\ \citenamefont
  {Novoselov}}]{Graphene_plasmonics-1}%
  \BibitemOpen
  \bibfield  {author} {\bibinfo {author} {\bibfnamefont {A.}~\bibnamefont
  {Grigorenko}}, \bibinfo {author} {\bibfnamefont {M.}~\bibnamefont {Polini}},
  \ and\ \bibinfo {author} {\bibfnamefont {K.}~\bibnamefont {Novoselov}},\
  }\href@noop {} {\bibfield  {journal} {\bibinfo  {journal} {Nat. Photonics}\
  }\textbf {\bibinfo {volume} {6}},\ \bibinfo {pages} {749} (\bibinfo {year}
  {2012})}\BibitemShut {NoStop}%
\bibitem [{\citenamefont {Koppens}\ \emph {et~al.}(2011)\citenamefont
  {Koppens}, \citenamefont {Chang},\ and\ \citenamefont {Garcia~de
  Abajo}}]{Graphene_plasmonics-2}%
  \BibitemOpen
  \bibfield  {author} {\bibinfo {author} {\bibfnamefont {F.~H.~L.}\
  \bibnamefont {Koppens}}, \bibinfo {author} {\bibfnamefont {D.~E.}\
  \bibnamefont {Chang}}, \ and\ \bibinfo {author} {\bibfnamefont {F.~J.}\
  \bibnamefont {Garcia~de Abajo}},\ }\href {\doibase 10.1021/nl201771h}
  {\bibfield  {journal} {\bibinfo  {journal} {Nano Lett.}\ }\textbf {\bibinfo
  {volume} {11}},\ \bibinfo {pages} {3370} (\bibinfo {year}
  {2011})}\BibitemShut {NoStop}%
\bibitem [{\citenamefont {Jablan}\ \emph {et~al.}(2009)\citenamefont {Jablan},
  \citenamefont {Buljan},\ and\ \citenamefont
  {Soljacik}}]{Graphene_plasmonics-3}%
  \BibitemOpen
  \bibfield  {author} {\bibinfo {author} {\bibfnamefont {M.}~\bibnamefont
  {Jablan}}, \bibinfo {author} {\bibfnamefont {H.}~\bibnamefont {Buljan}}, \
  and\ \bibinfo {author} {\bibfnamefont {M.}~\bibnamefont {Soljacik}},\ }\href
  {\doibase 10.1103/PhysRevB.80.245435} {\bibfield  {journal} {\bibinfo
  {journal} {Phys. Rev. B}\ }\textbf {\bibinfo {volume} {80}},\ \bibinfo
  {pages} {245435} (\bibinfo {year} {2009})}\BibitemShut {NoStop}%
\bibitem [{\citenamefont {Ryzhii}\ \emph {et~al.}(2007)\citenamefont {Ryzhii},
  \citenamefont {Satou},\ and\ \citenamefont {Otsuji}}]{Ryzhii-plasmons}%
  \BibitemOpen
  \bibfield  {author} {\bibinfo {author} {\bibfnamefont {V.}~\bibnamefont
  {Ryzhii}}, \bibinfo {author} {\bibfnamefont {A.}~\bibnamefont {Satou}}, \
  and\ \bibinfo {author} {\bibfnamefont {T.}~\bibnamefont {Otsuji}},\ }\href
  {\doibase 10.1063/1.2426904} {\bibfield  {journal} {\bibinfo  {journal} {J.
  Appl. Phys.}\ }\textbf {\bibinfo {volume} {101}},\ \bibinfo {eid} {024509}
  (\bibinfo {year} {2007})}\BibitemShut {NoStop}%
\bibitem [{\citenamefont {Hwang}\ and\ \citenamefont
  {Das~Sarma}(2007)}]{Das_Sarma_Plasmons}%
  \BibitemOpen
  \bibfield  {author} {\bibinfo {author} {\bibfnamefont {E.~H.}\ \bibnamefont
  {Hwang}}\ and\ \bibinfo {author} {\bibfnamefont {S.}~\bibnamefont
  {Das~Sarma}},\ }\href {\doibase 10.1103/PhysRevB.75.205418} {\bibfield
  {journal} {\bibinfo  {journal} {Phys. Rev. B}\ }\textbf {\bibinfo {volume}
  {75}},\ \bibinfo {pages} {205418} (\bibinfo {year} {2007})}\BibitemShut
  {NoStop}%
\bibitem [{\citenamefont {Mikhailov}\ and\ \citenamefont
  {Ziegler}(2007)}]{Mikhailov_new_mode}%
  \BibitemOpen
  \bibfield  {author} {\bibinfo {author} {\bibfnamefont {S.~A.}\ \bibnamefont
  {Mikhailov}}\ and\ \bibinfo {author} {\bibfnamefont {K.}~\bibnamefont
  {Ziegler}},\ }\href {\doibase 10.1103/PhysRevLett.99.016803} {\bibfield
  {journal} {\bibinfo  {journal} {Phys. Rev. Lett.}\ }\textbf {\bibinfo
  {volume} {99}},\ \bibinfo {pages} {016803} (\bibinfo {year}
  {2007})}\BibitemShut {NoStop}%
\bibitem [{\citenamefont {Svintsov}\ \emph {et~al.}(2012)\citenamefont
  {Svintsov}, \citenamefont {Vyurkov}, \citenamefont {Yurchenko}, \citenamefont
  {Otsuji},\ and\ \citenamefont {Ryzhii}}]{Our-hydrodynamic}%
  \BibitemOpen
  \bibfield  {author} {\bibinfo {author} {\bibfnamefont {D.}~\bibnamefont
  {Svintsov}}, \bibinfo {author} {\bibfnamefont {V.}~\bibnamefont {Vyurkov}},
  \bibinfo {author} {\bibfnamefont {S.}~\bibnamefont {Yurchenko}}, \bibinfo
  {author} {\bibfnamefont {T.}~\bibnamefont {Otsuji}}, \ and\ \bibinfo {author}
  {\bibfnamefont {V.}~\bibnamefont {Ryzhii}},\ }\href {\doibase
  10.1063/1.4705382} {\bibfield  {journal} {\bibinfo  {journal} {J. Appl.
  Phys.}\ }\textbf {\bibinfo {volume} {111}},\ \bibinfo {eid} {083715}
  (\bibinfo {year} {2012})}\BibitemShut {NoStop}%
\bibitem [{\citenamefont {Gangadharaiah}\ \emph {et~al.}(2008)\citenamefont
  {Gangadharaiah}, \citenamefont {Farid},\ and\ \citenamefont
  {Mishchenko}}]{New_plasmon_mode}%
  \BibitemOpen
  \bibfield  {author} {\bibinfo {author} {\bibfnamefont {S.}~\bibnamefont
  {Gangadharaiah}}, \bibinfo {author} {\bibfnamefont {A.~M.}\ \bibnamefont
  {Farid}}, \ and\ \bibinfo {author} {\bibfnamefont {E.~G.}\ \bibnamefont
  {Mishchenko}},\ }\href {\doibase 10.1103/PhysRevLett.100.166802} {\bibfield
  {journal} {\bibinfo  {journal} {Phys. Rev. Lett.}\ }\textbf {\bibinfo
  {volume} {100}},\ \bibinfo {pages} {166802} (\bibinfo {year}
  {2008})}\BibitemShut {NoStop}%
\bibitem [{\citenamefont {Mishchenko}\ \emph {et~al.}(2010)\citenamefont
  {Mishchenko}, \citenamefont {Shytov},\ and\ \citenamefont
  {Silvestrov}}]{Plasmons_pn}%
  \BibitemOpen
  \bibfield  {author} {\bibinfo {author} {\bibfnamefont {E.~G.}\ \bibnamefont
  {Mishchenko}}, \bibinfo {author} {\bibfnamefont {A.~V.}\ \bibnamefont
  {Shytov}}, \ and\ \bibinfo {author} {\bibfnamefont {P.~G.}\ \bibnamefont
  {Silvestrov}},\ }\href {\doibase 10.1103/PhysRevLett.104.156806} {\bibfield
  {journal} {\bibinfo  {journal} {Phys. Rev. Lett.}\ }\textbf {\bibinfo
  {volume} {104}},\ \bibinfo {pages} {156806} (\bibinfo {year}
  {2010})}\BibitemShut {NoStop}%
\bibitem [{\citenamefont {Tomadin}\ and\ \citenamefont
  {Polini}(2013)}]{Polini_FET}%
  \BibitemOpen
  \bibfield  {author} {\bibinfo {author} {\bibfnamefont {A.}~\bibnamefont
  {Tomadin}}\ and\ \bibinfo {author} {\bibfnamefont {M.}~\bibnamefont
  {Polini}},\ }\href {\doibase 10.1103/PhysRevB.88.205426} {\bibfield
  {journal} {\bibinfo  {journal} {Phys. Rev. B}\ }\textbf {\bibinfo {volume}
  {88}},\ \bibinfo {pages} {205426} (\bibinfo {year} {2013})}\BibitemShut
  {NoStop}%
\bibitem [{\citenamefont {Svintsov}\ \emph
  {et~al.}(2013{\natexlab{a}})\citenamefont {Svintsov}, \citenamefont
  {Vyurkov}, \citenamefont {Ryzhii},\ and\ \citenamefont {Otsuji}}]{Our_NLHD}%
  \BibitemOpen
  \bibfield  {author} {\bibinfo {author} {\bibfnamefont {D.}~\bibnamefont
  {Svintsov}}, \bibinfo {author} {\bibfnamefont {V.}~\bibnamefont {Vyurkov}},
  \bibinfo {author} {\bibfnamefont {V.}~\bibnamefont {Ryzhii}}, \ and\ \bibinfo
  {author} {\bibfnamefont {T.}~\bibnamefont {Otsuji}},\ }\href {\doibase
  10.1103/PhysRevB.88.245444} {\bibfield  {journal} {\bibinfo  {journal} {Phys.
  Rev. B}\ }\textbf {\bibinfo {volume} {88}},\ \bibinfo {pages} {245444}
  (\bibinfo {year} {2013}{\natexlab{a}})}\BibitemShut {NoStop}%
\bibitem [{\citenamefont {Ryzhii}\ \emph {et~al.}(2012)\citenamefont {Ryzhii},
  \citenamefont {Otsuji}, \citenamefont {Ryzhii}, \citenamefont {Leiman},
  \citenamefont {Yurchenko}, \citenamefont {Mitin},\ and\ \citenamefont
  {Shur}}]{Plasma-resonances-in-modulator}%
  \BibitemOpen
  \bibfield  {author} {\bibinfo {author} {\bibfnamefont {V.}~\bibnamefont
  {Ryzhii}}, \bibinfo {author} {\bibfnamefont {T.}~\bibnamefont {Otsuji}},
  \bibinfo {author} {\bibfnamefont {M.}~\bibnamefont {Ryzhii}}, \bibinfo
  {author} {\bibfnamefont {V.~G.}\ \bibnamefont {Leiman}}, \bibinfo {author}
  {\bibfnamefont {S.~O.}\ \bibnamefont {Yurchenko}}, \bibinfo {author}
  {\bibfnamefont {V.}~\bibnamefont {Mitin}}, \ and\ \bibinfo {author}
  {\bibfnamefont {M.~S.}\ \bibnamefont {Shur}},\ }\href {\doibase
  http://dx.doi.org/10.1063/1.4766814} {\bibfield  {journal} {\bibinfo
  {journal} {J. Appl. Phys.}\ }\textbf {\bibinfo {volume} {112}},\ \bibinfo
  {eid} {104507} (\bibinfo {year} {2012})}\BibitemShut {NoStop}%
\bibitem [{\citenamefont {Svintsov}\ \emph {et~al.}(2014)\citenamefont
  {Svintsov}, \citenamefont {Leiman}, \citenamefont {Ryzhii}, \citenamefont
  {Otsuji},\ and\ \citenamefont {Shur}}]{Our_NEMS}%
  \BibitemOpen
  \bibfield  {author} {\bibinfo {author} {\bibfnamefont {D.}~\bibnamefont
  {Svintsov}}, \bibinfo {author} {\bibfnamefont {V.~G.}\ \bibnamefont
  {Leiman}}, \bibinfo {author} {\bibfnamefont {V.}~\bibnamefont {Ryzhii}},
  \bibinfo {author} {\bibfnamefont {T.}~\bibnamefont {Otsuji}}, \ and\ \bibinfo
  {author} {\bibfnamefont {M.~S.}\ \bibnamefont {Shur}},\ }\href
  {http://stacks.iop.org/0022-3727/47/i=50/a=505105} {\bibfield  {journal}
  {\bibinfo  {journal} {J. Phys. D: Appl. Phys.}\ }\textbf {\bibinfo {volume}
  {47}},\ \bibinfo {pages} {505105} (\bibinfo {year} {2014})}\BibitemShut
  {NoStop}%
\bibitem [{\citenamefont {Fei}\ \emph {et~al.}(2012)\citenamefont {Fei},
  \citenamefont {Rodin}, \citenamefont {Andreev}, \citenamefont {Bao},
  \citenamefont {McLeod}, \citenamefont {Wagner}, \citenamefont {Zhang},
  \citenamefont {Zhao}, \citenamefont {Thiemens}, \citenamefont {Dominguez},
  \citenamefont {Fogler}, \citenamefont {Castro~Neto}, \citenamefont {Lau},
  \citenamefont {F.},\ and\ \citenamefont {Basov}}]{Fei_nano_imaging}%
  \BibitemOpen
  \bibfield  {author} {\bibinfo {author} {\bibfnamefont {Z.}~\bibnamefont
  {Fei}}, \bibinfo {author} {\bibfnamefont {A.}~\bibnamefont {Rodin}}, \bibinfo
  {author} {\bibfnamefont {G.}~\bibnamefont {Andreev}}, \bibinfo {author}
  {\bibfnamefont {W.}~\bibnamefont {Bao}}, \bibinfo {author} {\bibfnamefont
  {A.}~\bibnamefont {McLeod}}, \bibinfo {author} {\bibfnamefont
  {M.}~\bibnamefont {Wagner}}, \bibinfo {author} {\bibfnamefont
  {L.}~\bibnamefont {Zhang}}, \bibinfo {author} {\bibfnamefont
  {Z.}~\bibnamefont {Zhao}}, \bibinfo {author} {\bibfnamefont {M.}~\bibnamefont
  {Thiemens}}, \bibinfo {author} {\bibfnamefont {G.}~\bibnamefont {Dominguez}},
  \bibinfo {author} {\bibfnamefont {M.~M.}\ \bibnamefont {Fogler}}, \bibinfo
  {author} {\bibfnamefont {A.~H.}\ \bibnamefont {Castro~Neto}}, \bibinfo
  {author} {\bibfnamefont {C.~N.}\ \bibnamefont {Lau}}, \bibinfo {author}
  {\bibfnamefont {K.}~\bibnamefont {F.}}, \ and\ \bibinfo {author}
  {\bibfnamefont {D.~N.}\ \bibnamefont {Basov}},\ }\href@noop {} {\bibfield
  {journal} {\bibinfo  {journal} {Nature}\ }\textbf {\bibinfo {volume} {487}},\
  \bibinfo {pages} {82} (\bibinfo {year} {2012})}\BibitemShut {NoStop}%
\bibitem [{\citenamefont {Chen}\ \emph {et~al.}(2012)\citenamefont {Chen},
  \citenamefont {Badioli}, \citenamefont {Alonso-Gonz{\'a}lez}, \citenamefont
  {Thongrattanasiri}, \citenamefont {Huth}, \citenamefont {Osmond},
  \citenamefont {Spasenovi{\'c}}, \citenamefont {Centeno}, \citenamefont
  {Pesquera}, \citenamefont {Godignon}, \citenamefont {Elorza}, \citenamefont
  {Camara}, \citenamefont {de~Abajo}, \citenamefont {Hillenbrand},\ and\
  \citenamefont {Koppens}}]{Koppens_nano_imaging}%
  \BibitemOpen
  \bibfield  {author} {\bibinfo {author} {\bibfnamefont {J.}~\bibnamefont
  {Chen}}, \bibinfo {author} {\bibfnamefont {M.}~\bibnamefont {Badioli}},
  \bibinfo {author} {\bibfnamefont {P.}~\bibnamefont {Alonso-Gonz{\'a}lez}},
  \bibinfo {author} {\bibfnamefont {S.}~\bibnamefont {Thongrattanasiri}},
  \bibinfo {author} {\bibfnamefont {F.}~\bibnamefont {Huth}}, \bibinfo {author}
  {\bibfnamefont {J.}~\bibnamefont {Osmond}}, \bibinfo {author} {\bibfnamefont
  {M.}~\bibnamefont {Spasenovi{\'c}}}, \bibinfo {author} {\bibfnamefont
  {A.}~\bibnamefont {Centeno}}, \bibinfo {author} {\bibfnamefont
  {A.}~\bibnamefont {Pesquera}}, \bibinfo {author} {\bibfnamefont
  {P.}~\bibnamefont {Godignon}}, \bibinfo {author} {\bibfnamefont {A.~Z.}\
  \bibnamefont {Elorza}}, \bibinfo {author} {\bibfnamefont {N.}~\bibnamefont
  {Camara}}, \bibinfo {author} {\bibfnamefont {F.~J.~G.}\ \bibnamefont
  {de~Abajo}}, \bibinfo {author} {\bibfnamefont {R.}~\bibnamefont
  {Hillenbrand}}, \ and\ \bibinfo {author} {\bibfnamefont {F.~H.~L.}\
  \bibnamefont {Koppens}},\ }\href@noop {} {\bibfield  {journal} {\bibinfo
  {journal} {Nature}\ }\textbf {\bibinfo {volume} {487}},\ \bibinfo {pages}
  {77} (\bibinfo {year} {2012})}\BibitemShut {NoStop}%
\bibitem [{\citenamefont {Principi}\ \emph {et~al.}(2014)\citenamefont
  {Principi}, \citenamefont {Carrega}, \citenamefont {Lundeberg}, \citenamefont
  {Woessner}, \citenamefont {Koppens}, \citenamefont {Vignale},\ and\
  \citenamefont {Polini}}]{Principi_plasmon_loss_hBN}%
  \BibitemOpen
  \bibfield  {author} {\bibinfo {author} {\bibfnamefont {A.}~\bibnamefont
  {Principi}}, \bibinfo {author} {\bibfnamefont {M.}~\bibnamefont {Carrega}},
  \bibinfo {author} {\bibfnamefont {M.~B.}\ \bibnamefont {Lundeberg}}, \bibinfo
  {author} {\bibfnamefont {A.}~\bibnamefont {Woessner}}, \bibinfo {author}
  {\bibfnamefont {F.~H.~L.}\ \bibnamefont {Koppens}}, \bibinfo {author}
  {\bibfnamefont {G.}~\bibnamefont {Vignale}}, \ and\ \bibinfo {author}
  {\bibfnamefont {M.}~\bibnamefont {Polini}},\ }\href {\doibase
  10.1103/PhysRevB.90.165408} {\bibfield  {journal} {\bibinfo  {journal} {Phys.
  Rev. B}\ }\textbf {\bibinfo {volume} {90}},\ \bibinfo {pages} {165408}
  (\bibinfo {year} {2014})}\BibitemShut {NoStop}%
\bibitem [{\citenamefont {Woessner}\ \emph {et~al.}(2014)\citenamefont
  {Woessner}, \citenamefont {Lundeberg}, \citenamefont {Gao}, \citenamefont
  {Principi}, \citenamefont {Alonso-Gonz{\'a}lez}, \citenamefont {Carrega},
  \citenamefont {Watanabe}, \citenamefont {Taniguchi}, \citenamefont {Vignale},
  \citenamefont {Polini}, \citenamefont {Hone}, \citenamefont {Hillenbrand},\
  and\ \citenamefont {Koppens}}]{Koppens_nano_imaging_hBN}%
  \BibitemOpen
  \bibfield  {author} {\bibinfo {author} {\bibfnamefont {A.}~\bibnamefont
  {Woessner}}, \bibinfo {author} {\bibfnamefont {M.~B.}\ \bibnamefont
  {Lundeberg}}, \bibinfo {author} {\bibfnamefont {Y.}~\bibnamefont {Gao}},
  \bibinfo {author} {\bibfnamefont {A.}~\bibnamefont {Principi}}, \bibinfo
  {author} {\bibfnamefont {P.}~\bibnamefont {Alonso-Gonz{\'a}lez}}, \bibinfo
  {author} {\bibfnamefont {M.}~\bibnamefont {Carrega}}, \bibinfo {author}
  {\bibfnamefont {K.}~\bibnamefont {Watanabe}}, \bibinfo {author}
  {\bibfnamefont {T.}~\bibnamefont {Taniguchi}}, \bibinfo {author}
  {\bibfnamefont {G.}~\bibnamefont {Vignale}}, \bibinfo {author} {\bibfnamefont
  {M.}~\bibnamefont {Polini}}, \bibinfo {author} {\bibfnamefont
  {J.}~\bibnamefont {Hone}}, \bibinfo {author} {\bibfnamefont {R.}~\bibnamefont
  {Hillenbrand}}, \ and\ \bibinfo {author} {\bibfnamefont {F.~H.~L.}\
  \bibnamefont {Koppens}},\ }\href@noop {} {\bibfield  {journal} {\bibinfo
  {journal} {Nat. Mater.}\ } (\bibinfo {year} {2014})}\BibitemShut {NoStop}%
\bibitem [{\citenamefont {Mikhailov}(2013)}]{Mikhailov-THz}%
  \BibitemOpen
  \bibfield  {author} {\bibinfo {author} {\bibfnamefont {S.~A.}\ \bibnamefont
  {Mikhailov}},\ }\href {\doibase 10.1103/PhysRevB.87.115405} {\bibfield
  {journal} {\bibinfo  {journal} {Phys. Rev. B}\ }\textbf {\bibinfo {volume}
  {87}},\ \bibinfo {pages} {115405} (\bibinfo {year} {2013})}\BibitemShut
  {NoStop}%
\bibitem [{\citenamefont {Sensale-Rodriguez}(2013)}]{Berardi_APL}%
  \BibitemOpen
  \bibfield  {author} {\bibinfo {author} {\bibfnamefont {B.}~\bibnamefont
  {Sensale-Rodriguez}},\ }\href {\doibase http://dx.doi.org/10.1063/1.4821221}
  {\bibfield  {journal} {\bibinfo  {journal} {Appl. Phys. Lett.}\ }\textbf
  {\bibinfo {volume} {103}},\ \bibinfo {eid} {123109} (\bibinfo {year}
  {2013})}\BibitemShut {NoStop}%
\bibitem [{\citenamefont {Dubinov}\ \emph {et~al.}(2011)\citenamefont
  {Dubinov}, \citenamefont {Aleshkin}, \citenamefont {Mitin}, \citenamefont
  {Otsuji},\ and\ \citenamefont {Ryzhii}}]{Dubinov_JPCM}%
  \BibitemOpen
  \bibfield  {author} {\bibinfo {author} {\bibfnamefont {A.~A.}\ \bibnamefont
  {Dubinov}}, \bibinfo {author} {\bibfnamefont {V.~Y.}\ \bibnamefont
  {Aleshkin}}, \bibinfo {author} {\bibfnamefont {V.}~\bibnamefont {Mitin}},
  \bibinfo {author} {\bibfnamefont {T.}~\bibnamefont {Otsuji}}, \ and\ \bibinfo
  {author} {\bibfnamefont {V.}~\bibnamefont {Ryzhii}},\ }\href
  {http://stacks.iop.org/0953-8984/23/i=14/a=145302} {\bibfield  {journal}
  {\bibinfo  {journal} {J. Phys.: Cond. Mat.}\ }\textbf {\bibinfo {volume}
  {23}},\ \bibinfo {pages} {145302} (\bibinfo {year} {2011})}\BibitemShut
  {NoStop}%
\bibitem [{\citenamefont {Rana}(2008)}]{Rana_IEEE}%
  \BibitemOpen
  \bibfield  {author} {\bibinfo {author} {\bibfnamefont {F.}~\bibnamefont
  {Rana}},\ }\href {\doibase 10.1109/TNANO.2007.910334} {\bibfield  {journal}
  {\bibinfo  {journal} {IEEE T. Nanotechnol.}\ }\textbf {\bibinfo {volume}
  {7}},\ \bibinfo {pages} {91} (\bibinfo {year} {2008})}\BibitemShut {NoStop}%
\bibitem [{\citenamefont {Winnerl}\ \emph {et~al.}(2011)\citenamefont
  {Winnerl}, \citenamefont {Orlita}, \citenamefont {Plochocka}, \citenamefont
  {Kossacki}, \citenamefont {Potemski}, \citenamefont {Winzer}, \citenamefont
  {Malic}, \citenamefont {Knorr}, \citenamefont {Sprinkle}, \citenamefont
  {Berger}, \citenamefont {de~Heer}, \citenamefont {Schneider},\ and\
  \citenamefont {Helm}}]{Winnerl_PRL}%
  \BibitemOpen
  \bibfield  {author} {\bibinfo {author} {\bibfnamefont {S.}~\bibnamefont
  {Winnerl}}, \bibinfo {author} {\bibfnamefont {M.}~\bibnamefont {Orlita}},
  \bibinfo {author} {\bibfnamefont {P.}~\bibnamefont {Plochocka}}, \bibinfo
  {author} {\bibfnamefont {P.}~\bibnamefont {Kossacki}}, \bibinfo {author}
  {\bibfnamefont {M.}~\bibnamefont {Potemski}}, \bibinfo {author}
  {\bibfnamefont {T.}~\bibnamefont {Winzer}}, \bibinfo {author} {\bibfnamefont
  {E.}~\bibnamefont {Malic}}, \bibinfo {author} {\bibfnamefont
  {A.}~\bibnamefont {Knorr}}, \bibinfo {author} {\bibfnamefont
  {M.}~\bibnamefont {Sprinkle}}, \bibinfo {author} {\bibfnamefont
  {C.}~\bibnamefont {Berger}}, \bibinfo {author} {\bibfnamefont {W.~A.}\
  \bibnamefont {de~Heer}}, \bibinfo {author} {\bibfnamefont {H.}~\bibnamefont
  {Schneider}}, \ and\ \bibinfo {author} {\bibfnamefont {M.}~\bibnamefont
  {Helm}},\ }\href {\doibase 10.1103/PhysRevLett.107.237401} {\bibfield
  {journal} {\bibinfo  {journal} {Phys. Rev. Lett.}\ }\textbf {\bibinfo
  {volume} {107}},\ \bibinfo {pages} {237401} (\bibinfo {year}
  {2011})}\BibitemShut {NoStop}%
\bibitem [{\citenamefont {Gierz}\ \emph {et~al.}(2015)\citenamefont {Gierz},
  \citenamefont {Mitrano}, \citenamefont {Petersen}, \citenamefont {Cacho},
  \citenamefont {Turcu}, \citenamefont {Springate}, \citenamefont {St{\"o}hr},
  \citenamefont {K{\"o}hler}, \citenamefont {Starke},\ and\ \citenamefont
  {Cavalleri}}]{Gierz_JPCM}%
  \BibitemOpen
  \bibfield  {author} {\bibinfo {author} {\bibfnamefont {I.}~\bibnamefont
  {Gierz}}, \bibinfo {author} {\bibfnamefont {M.}~\bibnamefont {Mitrano}},
  \bibinfo {author} {\bibfnamefont {J.~C.}\ \bibnamefont {Petersen}}, \bibinfo
  {author} {\bibfnamefont {C.}~\bibnamefont {Cacho}}, \bibinfo {author}
  {\bibfnamefont {I.~C.~E.}\ \bibnamefont {Turcu}}, \bibinfo {author}
  {\bibfnamefont {E.}~\bibnamefont {Springate}}, \bibinfo {author}
  {\bibfnamefont {A.}~\bibnamefont {St{\"o}hr}}, \bibinfo {author}
  {\bibfnamefont {A.}~\bibnamefont {K{\"o}hler}}, \bibinfo {author}
  {\bibfnamefont {U.}~\bibnamefont {Starke}}, \ and\ \bibinfo {author}
  {\bibfnamefont {A.}~\bibnamefont {Cavalleri}},\ }\href
  {http://stacks.iop.org/0953-8984/27/i=16/a=164204} {\bibfield  {journal}
  {\bibinfo  {journal} {J. Phys.: Cond. Mat.}\ }\textbf {\bibinfo {volume}
  {27}},\ \bibinfo {pages} {164204} (\bibinfo {year} {2015})}\BibitemShut
  {NoStop}%
\bibitem [{\citenamefont {Lambe}\ and\ \citenamefont
  {McCarthy}(1976)}]{Lambe_PRL}%
  \BibitemOpen
  \bibfield  {author} {\bibinfo {author} {\bibfnamefont {J.}~\bibnamefont
  {Lambe}}\ and\ \bibinfo {author} {\bibfnamefont {S.~L.}\ \bibnamefont
  {McCarthy}},\ }\href {\doibase 10.1103/PhysRevLett.37.923} {\bibfield
  {journal} {\bibinfo  {journal} {Phys. Rev. Lett.}\ }\textbf {\bibinfo
  {volume} {37}},\ \bibinfo {pages} {923} (\bibinfo {year} {1976})}\BibitemShut
  {NoStop}%
\bibitem [{\citenamefont {Berndt}\ \emph {et~al.}(1991)\citenamefont {Berndt},
  \citenamefont {Gimzewski},\ and\ \citenamefont
  {Johansson}}]{Berndt-inelastic-tunneling-plasmons}%
  \BibitemOpen
  \bibfield  {author} {\bibinfo {author} {\bibfnamefont {R.}~\bibnamefont
  {Berndt}}, \bibinfo {author} {\bibfnamefont {J.~K.}\ \bibnamefont
  {Gimzewski}}, \ and\ \bibinfo {author} {\bibfnamefont {P.}~\bibnamefont
  {Johansson}},\ }\href {\doibase 10.1103/PhysRevLett.67.3796} {\bibfield
  {journal} {\bibinfo  {journal} {Phys. Rev. Lett.}\ }\textbf {\bibinfo
  {volume} {67}},\ \bibinfo {pages} {3796} (\bibinfo {year}
  {1991})}\BibitemShut {NoStop}%
\bibitem [{\citenamefont {Persson}\ and\ \citenamefont
  {Baratoff}(1992)}]{Persson-theory-tip-plasmon}%
  \BibitemOpen
  \bibfield  {author} {\bibinfo {author} {\bibfnamefont {B.~N.~J.}\
  \bibnamefont {Persson}}\ and\ \bibinfo {author} {\bibfnamefont
  {A.}~\bibnamefont {Baratoff}},\ }\href {\doibase 10.1103/PhysRevLett.68.3224}
  {\bibfield  {journal} {\bibinfo  {journal} {Phys. Rev. Lett.}\ }\textbf
  {\bibinfo {volume} {68}},\ \bibinfo {pages} {3224} (\bibinfo {year}
  {1992})}\BibitemShut {NoStop}%
\bibitem [{\citenamefont {Bharadwaj}\ \emph {et~al.}(2011)\citenamefont
  {Bharadwaj}, \citenamefont {Bouhelier},\ and\ \citenamefont
  {Novotny}}]{Novotny-tip-plasmon}%
  \BibitemOpen
  \bibfield  {author} {\bibinfo {author} {\bibfnamefont {P.}~\bibnamefont
  {Bharadwaj}}, \bibinfo {author} {\bibfnamefont {A.}~\bibnamefont
  {Bouhelier}}, \ and\ \bibinfo {author} {\bibfnamefont {L.}~\bibnamefont
  {Novotny}},\ }\href {\doibase 10.1103/PhysRevLett.106.226802} {\bibfield
  {journal} {\bibinfo  {journal} {Phys. Rev. Lett.}\ }\textbf {\bibinfo
  {volume} {106}},\ \bibinfo {pages} {226802} (\bibinfo {year}
  {2011})}\BibitemShut {NoStop}%
\bibitem [{\citenamefont {Kazarinov}\ and\ \citenamefont
  {Suris}(1972)}]{Kazarinov_Suris}%
  \BibitemOpen
  \bibfield  {author} {\bibinfo {author} {\bibfnamefont {R.}~\bibnamefont
  {Kazarinov}}\ and\ \bibinfo {author} {\bibfnamefont {R.}~\bibnamefont
  {Suris}},\ }\href@noop {} {\bibfield  {journal} {\bibinfo  {journal} {Sov.
  Phys. Semicond.}\ }\textbf {\bibinfo {volume} {6}},\ \bibinfo {pages} {148}
  (\bibinfo {year} {1972})}\BibitemShut {NoStop}%
\bibitem [{\citenamefont {Faist}\ \emph {et~al.}(1994)\citenamefont {Faist},
  \citenamefont {Capasso}, \citenamefont {Sivco}, \citenamefont {Sirtori},
  \citenamefont {Hutchinson},\ and\ \citenamefont {Cho}}]{Capasso_science}%
  \BibitemOpen
  \bibfield  {author} {\bibinfo {author} {\bibfnamefont {J.}~\bibnamefont
  {Faist}}, \bibinfo {author} {\bibfnamefont {F.}~\bibnamefont {Capasso}},
  \bibinfo {author} {\bibfnamefont {D.~L.}\ \bibnamefont {Sivco}}, \bibinfo
  {author} {\bibfnamefont {C.}~\bibnamefont {Sirtori}}, \bibinfo {author}
  {\bibfnamefont {A.~L.}\ \bibnamefont {Hutchinson}}, \ and\ \bibinfo {author}
  {\bibfnamefont {A.~Y.}\ \bibnamefont {Cho}},\ }\href@noop {} {\bibfield
  {journal} {\bibinfo  {journal} {Science}\ }\textbf {\bibinfo {volume}
  {264}},\ \bibinfo {pages} {553} (\bibinfo {year} {1994})}\BibitemShut
  {NoStop}%
\bibitem [{\citenamefont {Ryzhii}\ \emph {et~al.}(2013)\citenamefont {Ryzhii},
  \citenamefont {Dubinov}, \citenamefont {Aleshkin}, \citenamefont {Ryzhii},\
  and\ \citenamefont {Otsuji}}]{Ryzhii_DGL_laser}%
  \BibitemOpen
  \bibfield  {author} {\bibinfo {author} {\bibfnamefont {V.}~\bibnamefont
  {Ryzhii}}, \bibinfo {author} {\bibfnamefont {A.~A.}\ \bibnamefont {Dubinov}},
  \bibinfo {author} {\bibfnamefont {V.~Y.}\ \bibnamefont {Aleshkin}}, \bibinfo
  {author} {\bibfnamefont {M.}~\bibnamefont {Ryzhii}}, \ and\ \bibinfo {author}
  {\bibfnamefont {T.}~\bibnamefont {Otsuji}},\ }\href {\doibase
  http://dx.doi.org/10.1063/1.4826113} {\bibfield  {journal} {\bibinfo
  {journal} {Appl. Phys. Lett.}\ }\textbf {\bibinfo {volume} {103}},\ \bibinfo
  {eid} {163507} (\bibinfo {year} {2013})}\BibitemShut {NoStop}%
\bibitem [{\citenamefont {Fritz}\ \emph {et~al.}(2008)\citenamefont {Fritz},
  \citenamefont {Schmalian}, \citenamefont {M\"uller},\ and\ \citenamefont
  {Sachdev}}]{Fritz_PRB}%
  \BibitemOpen
  \bibfield  {author} {\bibinfo {author} {\bibfnamefont {L.}~\bibnamefont
  {Fritz}}, \bibinfo {author} {\bibfnamefont {J.}~\bibnamefont {Schmalian}},
  \bibinfo {author} {\bibfnamefont {M.}~\bibnamefont {M\"uller}}, \ and\
  \bibinfo {author} {\bibfnamefont {S.}~\bibnamefont {Sachdev}},\ }\href
  {\doibase 10.1103/PhysRevB.78.085416} {\bibfield  {journal} {\bibinfo
  {journal} {Phys. Rev. B}\ }\textbf {\bibinfo {volume} {78}},\ \bibinfo
  {pages} {085416} (\bibinfo {year} {2008})}\BibitemShut {NoStop}%
\bibitem [{\citenamefont {Mishchenko}\ \emph {et~al.}(2014)\citenamefont
  {Mishchenko}, \citenamefont {Tu}, \citenamefont {Cao}, \citenamefont
  {Gorbachev}, \citenamefont {Wallbank}, \citenamefont {Greenaway},
  \citenamefont {Morozov}, \citenamefont {Morozov}, \citenamefont {Zhu},
  \citenamefont {Wong}, \citenamefont {Withers}, \citenamefont {Woods},
  \citenamefont {Kim}, \citenamefont {Watanabe}, \citenamefont {Taniguchi},
  \citenamefont {Vdovin}, \citenamefont {Makarovsky}, \citenamefont {Fromhold},
  \citenamefont {Fal'ko}, \citenamefont {Geim}, \citenamefont {Eaves},\ and\
  \citenamefont {Novoselov}}]{Twist-controlled}%
  \BibitemOpen
  \bibfield  {author} {\bibinfo {author} {\bibfnamefont {A.}~\bibnamefont
  {Mishchenko}}, \bibinfo {author} {\bibfnamefont {J.}~\bibnamefont {Tu}},
  \bibinfo {author} {\bibfnamefont {Y.}~\bibnamefont {Cao}}, \bibinfo {author}
  {\bibfnamefont {R.}~\bibnamefont {Gorbachev}}, \bibinfo {author}
  {\bibfnamefont {J.}~\bibnamefont {Wallbank}}, \bibinfo {author}
  {\bibfnamefont {M.}~\bibnamefont {Greenaway}}, \bibinfo {author}
  {\bibfnamefont {V.}~\bibnamefont {Morozov}}, \bibinfo {author} {\bibfnamefont
  {S.}~\bibnamefont {Morozov}}, \bibinfo {author} {\bibfnamefont
  {M.}~\bibnamefont {Zhu}}, \bibinfo {author} {\bibfnamefont {S.}~\bibnamefont
  {Wong}}, \bibinfo {author} {\bibfnamefont {F.}~\bibnamefont {Withers}},
  \bibinfo {author} {\bibfnamefont {C.~R.}\ \bibnamefont {Woods}}, \bibinfo
  {author} {\bibfnamefont {Y.-J.}\ \bibnamefont {Kim}}, \bibinfo {author}
  {\bibfnamefont {K.}~\bibnamefont {Watanabe}}, \bibinfo {author}
  {\bibfnamefont {T.}~\bibnamefont {Taniguchi}}, \bibinfo {author}
  {\bibfnamefont {E.~E.}\ \bibnamefont {Vdovin}}, \bibinfo {author}
  {\bibfnamefont {O.}~\bibnamefont {Makarovsky}}, \bibinfo {author}
  {\bibfnamefont {T.}~\bibnamefont {Fromhold}}, \bibinfo {author}
  {\bibfnamefont {V.}~\bibnamefont {Fal'ko}}, \bibinfo {author} {\bibfnamefont
  {A.}~\bibnamefont {Geim}}, \bibinfo {author} {\bibfnamefont {L.}~\bibnamefont
  {Eaves}}, \ and\ \bibinfo {author} {\bibfnamefont {K.}~\bibnamefont
  {Novoselov}},\ }\href@noop {} {\bibfield  {journal} {\bibinfo  {journal}
  {Nat. Nanotechnol.}\ }\textbf {\bibinfo {volume} {9}},\ \bibinfo {pages}
  {808} (\bibinfo {year} {2014})}\BibitemShut {NoStop}%
\bibitem [{\citenamefont {Yadav}\ \emph {et~al.}(2015)\citenamefont {Yadav},
  \citenamefont {Tombet}, \citenamefont {Watanabe}, \citenamefont {Ryzhii},\
  and\ \citenamefont {Otsuji}}]{THz_emission_DGL_experim}%
  \BibitemOpen
  \bibfield  {author} {\bibinfo {author} {\bibfnamefont {D.}~\bibnamefont
  {Yadav}}, \bibinfo {author} {\bibfnamefont {S.}~\bibnamefont {Tombet}},
  \bibinfo {author} {\bibfnamefont {T.}~\bibnamefont {Watanabe}}, \bibinfo
  {author} {\bibfnamefont {V.}~\bibnamefont {Ryzhii}}, \ and\ \bibinfo {author}
  {\bibfnamefont {T.}~\bibnamefont {Otsuji}},\ }in\ \href {\doibase
  10.1109/DRC.2015.7175678} {\emph {\bibinfo {booktitle} {73rd Annual Device
  Research Conference (DRC)}}}\ (\bibinfo {year} {2015})\ pp.\ \bibinfo {pages}
  {271--272}\BibitemShut {NoStop}%
\bibitem [{\citenamefont {Hwang}\ and\ \citenamefont
  {Das~Sarma}(2009)}]{Hwang_PRB_2GL}%
  \BibitemOpen
  \bibfield  {author} {\bibinfo {author} {\bibfnamefont {E.~H.}\ \bibnamefont
  {Hwang}}\ and\ \bibinfo {author} {\bibfnamefont {S.}~\bibnamefont
  {Das~Sarma}},\ }\href {\doibase 10.1103/PhysRevB.80.205405} {\bibfield
  {journal} {\bibinfo  {journal} {Phys. Rev. B}\ }\textbf {\bibinfo {volume}
  {80}},\ \bibinfo {pages} {205405} (\bibinfo {year} {2009})}\BibitemShut
  {NoStop}%
\bibitem [{\citenamefont {Svintsov}\ \emph
  {et~al.}(2013{\natexlab{b}})\citenamefont {Svintsov}, \citenamefont
  {Vyurkov}, \citenamefont {Ryzhii},\ and\ \citenamefont
  {Otsuji}}]{Voltage_controlled}%
  \BibitemOpen
  \bibfield  {author} {\bibinfo {author} {\bibfnamefont {D.}~\bibnamefont
  {Svintsov}}, \bibinfo {author} {\bibfnamefont {V.}~\bibnamefont {Vyurkov}},
  \bibinfo {author} {\bibfnamefont {V.}~\bibnamefont {Ryzhii}}, \ and\ \bibinfo
  {author} {\bibfnamefont {T.}~\bibnamefont {Otsuji}},\ }\href {\doibase
  http://dx.doi.org/10.1063/1.4789818} {\bibfield  {journal} {\bibinfo
  {journal} {J. Appl. Phys.}\ }\textbf {\bibinfo {volume} {113}},\ \bibinfo
  {eid} {053701} (\bibinfo {year} {2013}{\natexlab{b}})}\BibitemShut {NoStop}%
\bibitem [{\citenamefont {Das~Sarma}\ and\ \citenamefont
  {Hwang}(1998)}]{DasSarma-PRL-tunnel-plasmon}%
  \BibitemOpen
  \bibfield  {author} {\bibinfo {author} {\bibfnamefont {S.}~\bibnamefont
  {Das~Sarma}}\ and\ \bibinfo {author} {\bibfnamefont {E.~H.}\ \bibnamefont
  {Hwang}},\ }\href {\doibase 10.1103/PhysRevLett.81.4216} {\bibfield
  {journal} {\bibinfo  {journal} {Phys. Rev. Lett.}\ }\textbf {\bibinfo
  {volume} {81}},\ \bibinfo {pages} {4216} (\bibinfo {year}
  {1998})}\BibitemShut {NoStop}%
\bibitem [{\citenamefont {Ryzhii}\ and\ \citenamefont
  {Shur}(2001)}]{Ryzhii_Shur_JJAP}%
  \BibitemOpen
  \bibfield  {author} {\bibinfo {author} {\bibfnamefont {V.}~\bibnamefont
  {Ryzhii}}\ and\ \bibinfo {author} {\bibfnamefont {M.}~\bibnamefont {Shur}},\
  }\href {http://stacks.iop.org/1347-4065/40/i=2R/a=546} {\bibfield  {journal}
  {\bibinfo  {journal} {Jpn. J. Appl. Phys.}\ }\textbf {\bibinfo {volume}
  {40}},\ \bibinfo {pages} {546} (\bibinfo {year} {2001})}\BibitemShut
  {NoStop}%
\bibitem [{\citenamefont {Brey}(2014)}]{Brey_PRA}%
  \BibitemOpen
  \bibfield  {author} {\bibinfo {author} {\bibfnamefont {L.}~\bibnamefont
  {Brey}},\ }\href {\doibase 10.1103/PhysRevApplied.2.014003} {\bibfield
  {journal} {\bibinfo  {journal} {Phys. Rev. Applied}\ }\textbf {\bibinfo
  {volume} {2}},\ \bibinfo {pages} {014003} (\bibinfo {year}
  {2014})}\BibitemShut {NoStop}%
\bibitem [{\citenamefont {Vasko}(2013)}]{Vasko_PRB}%
  \BibitemOpen
  \bibfield  {author} {\bibinfo {author} {\bibfnamefont {F.~T.}\ \bibnamefont
  {Vasko}},\ }\href {\doibase 10.1103/PhysRevB.87.075424} {\bibfield  {journal}
  {\bibinfo  {journal} {Phys. Rev. B}\ }\textbf {\bibinfo {volume} {87}},\
  \bibinfo {pages} {075424} (\bibinfo {year} {2013})}\BibitemShut {NoStop}%
\bibitem [{\citenamefont {Bistritzer}\ and\ \citenamefont
  {MacDonald}(2011)}]{Twisted_GBL}%
  \BibitemOpen
  \bibfield  {author} {\bibinfo {author} {\bibfnamefont {R.}~\bibnamefont
  {Bistritzer}}\ and\ \bibinfo {author} {\bibfnamefont {A.~H.}\ \bibnamefont
  {MacDonald}},\ }\href@noop {} {\bibfield  {journal} {\bibinfo  {journal}
  {Proc. Nat. Acad. Sci.}\ }\textbf {\bibinfo {volume} {108}},\ \bibinfo
  {pages} {12233} (\bibinfo {year} {2011})}\BibitemShut {NoStop}%
\bibitem [{\citenamefont {Vasko}\ and\ \citenamefont
  {Kuznetsov}(2012)}]{vasko_book}%
  \BibitemOpen
  \bibfield  {author} {\bibinfo {author} {\bibfnamefont {F.~T.}\ \bibnamefont
  {Vasko}}\ and\ \bibinfo {author} {\bibfnamefont {A.~V.}\ \bibnamefont
  {Kuznetsov}},\ }\href@noop {} {\emph {\bibinfo {title} {Electronic states and
  optical transitions in semiconductor heterostructures}}}\ (\bibinfo
  {publisher} {Springer Science \& Business Media},\ \bibinfo {year}
  {2012})\BibitemShut {NoStop}%
\bibitem [{\citenamefont {Shi}\ \emph {et~al.}(2013)\citenamefont {Shi},
  \citenamefont {Pan}, \citenamefont {Zhang},\ and\ \citenamefont
  {Yakobson}}]{Band_parameters_MOS2}%
  \BibitemOpen
  \bibfield  {author} {\bibinfo {author} {\bibfnamefont {H.}~\bibnamefont
  {Shi}}, \bibinfo {author} {\bibfnamefont {H.}~\bibnamefont {Pan}}, \bibinfo
  {author} {\bibfnamefont {Y.-W.}\ \bibnamefont {Zhang}}, \ and\ \bibinfo
  {author} {\bibfnamefont {B.~I.}\ \bibnamefont {Yakobson}},\ }\href {\doibase
  10.1103/PhysRevB.87.155304} {\bibfield  {journal} {\bibinfo  {journal} {Phys.
  Rev. B}\ }\textbf {\bibinfo {volume} {87}},\ \bibinfo {pages} {155304}
  (\bibinfo {year} {2013})}\BibitemShut {NoStop}%
\bibitem [{\citenamefont {Bolotin}\ \emph {et~al.}(2008)\citenamefont
  {Bolotin}, \citenamefont {Sikes}, \citenamefont {Hone}, \citenamefont
  {Stormer},\ and\ \citenamefont {Kim}}]{PRL_Bolotin}%
  \BibitemOpen
  \bibfield  {author} {\bibinfo {author} {\bibfnamefont {K.~I.}\ \bibnamefont
  {Bolotin}}, \bibinfo {author} {\bibfnamefont {K.~J.}\ \bibnamefont {Sikes}},
  \bibinfo {author} {\bibfnamefont {J.}~\bibnamefont {Hone}}, \bibinfo {author}
  {\bibfnamefont {H.~L.}\ \bibnamefont {Stormer}}, \ and\ \bibinfo {author}
  {\bibfnamefont {P.}~\bibnamefont {Kim}},\ }\href {\doibase
  10.1103/PhysRevLett.101.096802} {\bibfield  {journal} {\bibinfo  {journal}
  {Phys. Rev. Lett.}\ }\textbf {\bibinfo {volume} {101}},\ \bibinfo {pages}
  {096802} (\bibinfo {year} {2008})}\BibitemShut {NoStop}%
\bibitem [{\citenamefont {Vasko}\ and\ \citenamefont
  {Ryzhii}(2007)}]{Vasko-Ryzhii}%
  \BibitemOpen
  \bibfield  {author} {\bibinfo {author} {\bibfnamefont {F.~T.}\ \bibnamefont
  {Vasko}}\ and\ \bibinfo {author} {\bibfnamefont {V.}~\bibnamefont {Ryzhii}},\
  }\href {\doibase 10.1103/PhysRevB.76.233404} {\bibfield  {journal} {\bibinfo
  {journal} {Phys. Rev. B}\ }\textbf {\bibinfo {volume} {76}},\ \bibinfo
  {pages} {233404} (\bibinfo {year} {2007})}\BibitemShut {NoStop}%
\bibitem [{\citenamefont {Falkovsky}\ and\ \citenamefont
  {Varlamov}(2007)}]{Falkovsky-Varlamov}%
  \BibitemOpen
  \bibfield  {author} {\bibinfo {author} {\bibfnamefont {L.~A.}\ \bibnamefont
  {Falkovsky}}\ and\ \bibinfo {author} {\bibfnamefont {A.~A.}\ \bibnamefont
  {Varlamov}},\ }\href {\doibase 10.1140/epjb/e2007-00142-3} {\bibfield
  {journal} {\bibinfo  {journal} {Eur. Phys. J. B}\ }\textbf {\bibinfo {volume}
  {56}},\ \bibinfo {pages} {281} (\bibinfo {year} {2007})}\BibitemShut
  {NoStop}%
\bibitem [{\citenamefont {Mermin}(1970)}]{Mermin-Lindhard_dielectric_Function}%
  \BibitemOpen
  \bibfield  {author} {\bibinfo {author} {\bibfnamefont {N.~D.}\ \bibnamefont
  {Mermin}},\ }\href {\doibase 10.1103/PhysRevB.1.2362} {\bibfield  {journal}
  {\bibinfo  {journal} {Phys. Rev. B}\ }\textbf {\bibinfo {volume} {1}},\
  \bibinfo {pages} {2362} (\bibinfo {year} {1970})}\BibitemShut {NoStop}%
\bibitem [{\citenamefont {Bhatnagar}\ \emph {et~al.}(1954)\citenamefont
  {Bhatnagar}, \citenamefont {Gross},\ and\ \citenamefont
  {Krook}}]{BGK-collisions}%
  \BibitemOpen
  \bibfield  {author} {\bibinfo {author} {\bibfnamefont {P.~L.}\ \bibnamefont
  {Bhatnagar}}, \bibinfo {author} {\bibfnamefont {E.~P.}\ \bibnamefont
  {Gross}}, \ and\ \bibinfo {author} {\bibfnamefont {M.}~\bibnamefont
  {Krook}},\ }\href {\doibase 10.1103/PhysRev.94.511} {\bibfield  {journal}
  {\bibinfo  {journal} {Phys. Rev.}\ }\textbf {\bibinfo {volume} {94}},\
  \bibinfo {pages} {511} (\bibinfo {year} {1954})}\BibitemShut {NoStop}%
\end{thebibliography}%

\end{document}